\documentclass[preprint,11]{aastex}
\usepackage{verbatim,float}
\usepackage{times}
\usepackage{amsmath}
\usepackage{graphics}
\usepackage{epsfig}
\usepackage{psfig}

\begin{document}

\shorttitle{Spectroscopy of NGC 1399 globular clusters}
\shortauthors{Richtler et al.}

\title{The Globular Cluster System of NGC 1399. II. Kinematics of
a Large Sample of Globular Clusters \altaffilmark{9} }

\author{T. Richtler \altaffilmark{1}, B. Dirsch \altaffilmark{1}, K. Gebhardt
 \altaffilmark{2}, D. Geisler \altaffilmark{1}, M. Hilker \altaffilmark{3},
M.V. Alonso \altaffilmark{7}, J.C. Forte \altaffilmark{6}, E.K. Grebel \altaffilmark{8}, L. Infante \altaffilmark{4}, S. Larsen \altaffilmark{5}, D. Minniti \altaffilmark{4}, M. Rejkuba \altaffilmark{5}}

\altaffiltext{1}{Universidad de Concepci\'on, Departamento de F\'{\i}sica,
                Casilla 160-C, Concepci\'on, Chile}
\altaffiltext{2}{Department of Astronomy, University of Texas at Austin, USA}
\altaffiltext{3}{Astronomisches Institut der Universit\"at Bonn, Auf dem H\"ugel 71, 53121 Bonn, Germany}
\altaffiltext{4}{Departamento de Astronom\'{\i}a y Astrof\'{\i}sica, P. Universidad Cat\'olica, Vicu\~na Mackenna 4860, Santiago 22, Chile}
\altaffiltext{5}{European Southern Observatory, Garching, Karl-Schwarzschild-Str. 2,
Germany}
\altaffiltext{6}{Universidad Nacional de La Plata, Paseo del Bosque S/N, 1900-La Plata, Argentina}
\altaffiltext{7}{Observatorio Astr\'onomico de
C\'ordoba,  Laprida  854, C\'ordoba, 5000, Argentina and CONICET, Argentina\\
CNRS UMR 5572, Observatoire Midi-Pyr\'en\'ees, 14 Avenue E. Belin, 31400
Toulouse, France}
\altaffiltext{8}{Astronomisches Institut der Universit\"at Basel, Venusstrasse 7, CH-4102, Binningen, Switzerland}

\altaffiltext{9}{Based on observations collected at the European Southern
Observatory, Cerro Paranal, Chile; ESO program 66.B-0393.}
\email{tom@coma.cfm.udec.cl}
\begin{abstract}

We study the kinematics and dynamics of the globular cluster system
of NGC 1399, the central galaxy of the Fornax cluster. The
observational data consists of medium resolution spectra, obtained at the Very Large Telescope
with FORS2 and the Mask Exchange Unit (MXU). Our sample comprises 468
radial velocities in the magnitude range $20 < m_R < 23$. This is the largest sample of
globular cluster velocities around any galaxy obtained so far. 
Typical velocity uncertainties are
50 km/s, significantly improving on earlier samples.
 The radial range is $2 \arcmin < r < 9 \arcmin$, corresponding to  11 kpc to
50 kpc of galactocentric distance.

The shape of the velocity distribution of the sample is compatible
with being a Gaussian distribution. However, under moderate error selection,
a slight asymmetry is
visible between high and low radial velocities. We find bright clusters
with radial velocities below 800 km/s, while they are not found at the corresponding high velocity side above 2000 km/s. There is the possibility that unbound
clusters and/or objects in the foreground contaminate the NGC 1399 cluster
sample.   Under strong error selection, practically no objects are found with
velocities lower than 800 km/s or higher than 2000 km/s. 
Since the extreme velocities influence
the velocity dispersion considerably, uncertainty regarding the exact value
of the dispersion remains. 
With the above velocity limits, we derive a projected velocity dispersion for the total sample of 274$\pm$9 km/s which
within the uncertainties remains constant  over the entire radial range.  
Without any velocity restriction, it increases to 325 km/s.

Guided by the bimodal color distribution of clusters, we distinguish between
red clusters (C-R $>$1.6) and blue clusters (C-R $<$1.6), and find 
velocity dispersions for these groups of 255 $\pm$13 km/s and 291$\pm$14 km/s,
respectively, again radially constant. 

Any possible rotation of either of these cluster populations is below
 the detection limit with the exception of a weak signature of rotation for the blue
clusters more distant than 6$\arcmin$. 

Spherical models point to  a circular velocity of $419\pm30$ km/s, assuming 
isotropy for the red clusters. This value is constant
out to 40 kpc.  
The inferred dark halo potential can be well represented by a logarithmic
potential. Also a halo of the NFW type fits well to the observations. 
 The orbital structure of the clusters can only be weakly constrained. It is consistent
with isotropy for the red clusters and a slight tangential bias for the blue clusters.

Some mass profiles derived from X-ray analyses do not agree with a constant
circular velocity within our radial range, irrespective of its exact value.
Interpreting the  extreme low radial velocities as space velocities of bound clusters
near their pericentric distances would require an extension of the cluster system of at least 200 kpc. 

Implications for formation scenarios of the cluster system are briefly commented on.

\end{abstract}



\section{Introduction}
\subsection{General remarks and introduction to the relevant literature}

To measure  the mass profiles of early-type galaxies is not as straightforward
as in the case of spiral galaxies, where a more or less orderly rotating
disk allows a comparatively easy derivation of the rotation curve  
(see Sofue \& Rubin \cite{sofue2001} for a review).  
The inclination of a stellar or HI disk, which in principle can be
measured by the axis ratio, leads to
the deprojection of the observed radial velocities and almost circular
orbits directly yield the circular velocities $v_c$:
\begin{center}
\begin{equation}
 v_c^2 = \frac{G \cdot M(r)}{r}, 
\end{equation}
\end{center}
where $G$ is the constant of gravitation, $r$ the galactocentric distance,
and $M(r)$ the mass within $r$.
Such favourable conditions do not exist for elliptical galaxies. 
An elliptical
galaxy can in reality be axisymmetric when appearing round in projection. 
Further complications can result from non-symmetric structure due to
recent merger events. Moreover, the 
distribution of stellar orbits is a priori unknown (see de Zeeuw \cite{dezeeuw1994} for
an introduction to the dynamics of elliptical galaxies).
Therefore, much effort has been devoted to develop 
 more sophisticated stellar dynamical methods in order to derive $M(r)$ for
elliptical galaxies from the analysis of the line-of-sight velocity
distribution (e.g. Dejonghe \cite{dejonghe1987}, van der Marel \& Franx \cite{marel1993}, Gerhard \cite{gerhard1993}, Bender et al. \cite{bender1994}, Rix et al. 
\cite{rix1997}, Gebhardt et al. \cite{gebhardt2000}). 

One of the most important objectives is  to prove the 
existence of dark matter halos and to investigate their structure.  
For a few X-ray bright ellipticals, this has been possible by assuming
hydrodynamical equilibrium for the hot gaseous halo (e.g. Fabricant \&
Gorenstein \cite{fabricant1983}, Nulsen \& B\"ohringer \cite{nulsen1995} (M87), Matilsky et al. \cite{matilsky1985} (NGC 4696),
Foreman et al. \cite{foreman1985}(13 galaxies), Irwin \& Sarazin \cite[1996]{irwin1996}
 (NGC4472), Mushotzky et al. \cite{musho1994} (NGC 4636),
Jones et al. \cite{jones1997} (NGC 1399), Ikebe et al. \cite{ikebe1996} 
(NGC 1399), Paolillo et al. \cite{pao2002}(NGC 1399)). Given the
uncertainties and limited applicability of this method (e.g. Buote \cite{buote2000}, Buote \cite{buote2002}, Paolillo et al. \cite{pao2002}), 
stellar dynamical approaches are very desirable. 
One faces a further problem in the case of elliptical galaxies:
while easily observable objects for determining the rotation curves of spirals sometimes extend to large galactocentric
radii (larger than 50 kpc), the rapidly declining brightness of the light
profile of ellipticals rarely allows spectroscopic work beyond 1.5 effective
radii. This is a distance where the effect of a dark halo is only just becoming
visible (Kronawitter et al. \cite{krona2000}, Gerhard et al. \cite{gerhard2001}). 
However, Kelson et al. \cite{kelson2002}, using the Keck telescope, recently achieved
 measurement of  
the velocity dispersion of the stellar population of NGC 6166 out to 60 kpc.

To extend stellar dynamical investigations to large galactocentric distances, one 
normally  depends
on the use of individual dynamical probes in the halos of elliptical galaxies.
These probes can be planetary nebulae (PNe) and/or globular clusters (GCs).
 The observational difficulty is that these objects are faint. There have been some attempts
in the past to exploit this kind of dynamical information with 4-m class
telescopes (Arnaboldi et al. \cite{arna1994} (NGC 1399; PNe), Grillmair et al.
 \cite{grillmair1994} (NGC 1399; GCs); Mould et al. \cite{mould1990} (M87; GCs)) but the sample sizes were small and the accuracy of the
radial velocities was low.   
However, a larger sample of about 100 radial velocities of GCs has been observed by Zepf et
al. \cite{zepf2000} for NGC 4472. The advent of large telescopes brought better possibilities
both in terms of number and accuracy. A sample of about 550 PNe-velocities has been
presented by M\'endez et al. \cite{mendez2000} for the flattened elliptical galaxy NGC 4607,
using the Very Large Telescope.
For NGC 4472  C\^ot\'e et al. \cite{cote2003} considerably augmented the sample of GC-velocities to
260 GCs, using the Keck telescope. 

The demands for a dynamically meaningful sample are severe: numerical simulations have shown
that several hundred, perhaps even thousand, radial velocities are required  if one intends to fit the potential and the phase space distribution simultaneously (e.g. Merritt \& Tremblay
\cite{merritt1993}).
 The
potential and the orbital properties of dynamical probes are
degenerate with respect to  radial velocities, so one has
to make specific assumptions concerning symmetry and isotropy/anisotropy if
one wants to use the observed velocity dispersions to infer the mass profile
with a smaller sample. 

M87, the most prominent GC-rich elliptical galaxy in the northern hemisphere, 
has in recent years become the target for ambitious studies regarding its
globular cluster system (GCS). Cohen \& Ryzhov \cite{cohen1997} and
Cohen \cite{cohen2000} observed about 200 radial velocities, using
the Keck telescope. 
This sample has been improved and enlarged to about 300 velocities by C\^ot\'e et al. \cite{cote2001} and  Hanes et al. \cite{hanes2001}, but even this large dataset has rather been 
 used to
explore the kinematics of the globular cluster system in relation to its
population properties than  to derive M(r).

In the southern hemisphere, the technological preconditions for such a demanding
project  have been considerably improved by the advent of sophisticated
 spectroscopic equipment at  large telescopes, which allows the observations of 
hundreds of velocities with satisfactory accuracy within a reasonable observing
time. 
The prime natural target is NGC 1399 in the Fornax cluster, given its proximity
and populous GCS.


\subsection{NGC 1399}
NGC 1399 is a particularly attractive galaxy with regards to the study of dark matter
halos. It is the brightest elliptical galaxy in the nearest galaxy cluster
in the Southern hemisphere, the Fornax cluster (e.g. Drinkwater et al. \cite{drink2001} 
and references therein). It has a distinguished 
location near the center of this  cluster. In the literature, it has
been sometimes labelled as the central cD galaxy. It is not within the
 scope of this paper 
to expand on these characteristics, but see Dirsch et al. \cite{dirsch2003}
(hereafter Paper I), who
present the first wide-field CCD study, for a discussion
as to why this label is questionable.

Since the distance is one of the most fundamental parameters in all
analyses, we include a brief summary of what is known of the distance
of NGC 1399 and the Fornax cluster.
A compilation of older work on the Fornax cluster using  several
methods can be found in Richtler et al. \cite{richtler1999}, resulting in a mean distance
modulus of 31.36$\pm$0.06.
Ferrarese et al. \cite{ferra2000} calibrated several methods and compiled distances
for 10 early-type Fornax galaxies, giving 31.44 (with a very small
formal error) as the average. They list NGC 1399 with 31.31$\pm$0.09.
The catalog of surface brightness fluctuation measurements by Tonry et al. \cite{tonry2001} includes 16 galaxies in the
core of the Fornax cluster, for which one derives a mean modulus of
31.50$\pm$0.05 and NGC 1399 is given as 31.50$\pm$0.16.
 Note that none of the distance uncertainties  include the
uncertainty in the absolute scale, which is around $\pm$0.16 mag, 
mainly due to the LMC distance  uncertainty.
This gives some freedom in the choice of the distance modulus. A value of 31.40
lies within the uncertainty interval of all measurements made so far and
we will adopt it here. The corresponding distance then is 19 Mpc.

NGC 1399 offers a wealth of GCs. The total number from recent wide-field
photometry is about 7000 (Paper I). 
The central galaxies in galaxy clusters often have the
highest specific frequencies of GCSs occuring among elliptical galaxies. The
specific frequency is defined as $S_N = N \cdot 10^{0.4 \cdot (M_V+15)}$,
where $N$ is the total number of GCs and $M_V$ the absolute V-magnitude of the
host galaxy (see Elmegreen \cite{elm2000} for a review). The classical value
for NGC 1399 is 10 or even larger (e.g. Wagner et al. \cite{wagner1991}, Bridges
et al. \cite{bridges1991}). However, several recent studies (Paper I and Ostrov et al. \cite{ostrov1998})  derive a value of
about 6. This change is mainly due to a better account of the total luminosity
of NGC 1399, as has been also mentioned by McLaughlin \cite{mac1999}. Although still  high for a giant elliptical,  there are also non-central
 galaxies like NGC 4728 
(Harris \& van den Bergh \cite{harris1981}), NGC 4636 (Kissler-Patig et al. \cite{kissler1994}),
 or IC 4051 (Woodworth \& Harris \cite{wood2000}) which reportedly  reach or even surpass  this
value.  

Dynamical studies of the stellar body of NGC 1399 have been performed
by Bicknell et al. \cite{bicknell1989}, Graham et al. \cite{graham1998}, Saglia et al. \cite{saglia2000}, and
Kronawitter et al. \cite{krona2000}.
Regarding the cluster system of NGC 1399, the situation was still unsatisfactory.
 Grillmair et al. \cite{grillmair1994}
obtained 46 velocities for GCs in NGC 1399, Kissler-Patig et al.
 \cite{kissler1999} augmented this sample to 76.
Spectroscopy for smaller samples of GCs, with the goal of measuring ages and
abundances rather than studying the dynamics, has been published by Mieske et al. \cite{mieske2002} and Forbes et al. \cite{forbes2001}.   

This  paper is the second in a series on the cluster system of NGC 1399.
Details of the data reduction  and the  dataset is presented in a parallel paper
 (Dirsch et al. \cite{dirsch2003b}, hereafter Paper III). New wide-field photometry in 
Washington C and Kron-Cousins R, from which we
selected the cluster candidates and derived the properties of the cluster
system needed in dynamical studies, is found in Paper I. These properties are the
 deprojected luminosity profile of NGC 1399 in R which has been
determined to
\begin{equation}
 [L_{\sun}/pc^3] = 101\cdot (1+r/221\mathrm{pc})^{-2.85}
\end{equation}
and the power-law exponents for the deprojected number density profiles
of blue ($1.2 < C-R < 1.6$) and red clusters ($1.6< C-R < 2.2$), which are
$-1.8 \pm 0.1$ and $-2.64 \pm 0.1$, respectively. 

The scope of the present paper is to show those results emerging from the
dataset which are possible to derive without
sophisticated dynamical modelling, which is left to a forthcoming paper
(Gebhardt et al. \cite{gebhardt2004}). We briefly present the observations (Section 2), present the morphology of the velocity distribution, investigate the rotation properties,
the observed (projected) velocity dispersions and their radial dependence, their significance for different
cluster populations (all in Section 3), interprete them in the context of a simple
spherical model (Section 4), and discuss the results in comparison with the literature.
Finally, we  comment on formation scenarios of the cluster system 
(Section 5). Section 6 contains the conclusions and the
outlook for further work.

\section{Observations and velocity measurements}

Paper III provides the details
of the slit mask preparation, the observations and reduction, and the database. Here we
only give a summary of the observations and the radial velocity 
measurements.

\subsection{Observations}
Paper I provides the candidate selection based on
wide-field photometry in Washington C and Kron-Cousins R obtained with  the CTIO
4m MOSAIC system.
The spectroscopic observations have been performed in the period 11/30-12/2\, 2000 at the Very
Large Telescope of the European Southern Observatory at Cerro Paranal with
Unit Telescope 2 (Kueyen). The instrument was the focal reducer FORS2 equipped
with  the Mask Exchange Unit (MXU), which allows the use of
up to ten slit masks during the night (for details see http://www.eso.org/instruments/fors/userman/). The grism in use was, for all but one mask, 600B, giving
a spectral resolution of about 2.5 {\AA}, based on the widths of the arc calibration lines. For one mask, we used grism 300V, which gave a resolution of about 5
{\AA}. The wavelength calibration was based on He-Ar lamp spectra.
 The spectral range covered about 2000 {\AA}. However, depending on the position
of a given slit in the mask, the limiting wavelengths can be as
short as 3500 {\AA} and as long as
 6500 {\AA}. In most cases, the
grism efficiency degraded the signal-to-noise shortwards of 3800 {\AA} drastically,
 so this region could not be used.

We exposed 13 masks (exposure times were either 1$\times$ or 2$\times$45 min per mask).
The FWHM of objects along the slit ranged from 0.6$\arcsec$ to 0.9$\arcsec$.
The total number of spectra obtained is 1462. This sample is composed of 531 sky
 spectra, 512 spectra of cluster candidates, 190 spectra of point sources of
unknown nature at the time of mask preparation (stars, unresolved
galaxies or clusters not matching the selection criteria), 176 galaxies, and
53 ``bright'' objects, mainly stars needed to adjust the mask.
 Since some objects (about 80) have been observed in
two different masks in order to assess the errors, the total number of objects is smaller by this number than
the number of spectra. We anticipate here that the final sample of cluster velocities
comprises  468 objects.

\subsection{Velocities}
We used two different techniques to measure the radial velocities: a 
cross-correlation technique and direct measurements of line positions.\\ 
For cross-correlation, we needed a template with a high S/N and a spectrum 
resembling that of a globular cluster. NGC 1399 itself is not suitable due
to its high velocity dispersion. We therefore chose NGC 1396 as a template,
which has a high S/N (about 30) and, as a dwarf elliptical  galaxy, has a
 spectrum like a metal-rich globular cluster. 
 Regarding line-measurements,
we defined a set of about 20 easily identifiable features (for which only
a few remained in
faint spectra) and used the IRAF task ``rvidlines'' for
measuring the radial velocities. In some cases, it was only possible to get
a velocity by cross-correlation and not by lines.  
For a more detailed discussion of the errors,
we refer the reader to Paper III.
Fig. \ref{errorplot} plots our uncertainties against the cluster magnitude.
For comparison we also include the M87 sample from Hanes et al.
 \cite{hanes2001}.

\section{Results}

\subsection{Total numbers, projected distribution, and absolute velocities}

The total number of clusters for which we have velocities is 508. However,
there are 40 objects for which we could not find a correlation velocity and
49 objects for which we could not derive a velocity by line positions due
to low S/N. In
the following, we prefer to use only the correlation velocities for reasons of homogeneity and because they have smaller uncertainties.
However, in the case of low S/N, the uncertainty may be underestimated, since
we used the width of the correlation peak to estimate the error.

For our template galaxy NGC 1396,
we derive a heliocentric radial velocity of 815 $\pm$ 8 km/s. This
is marginally in agreement with the value of 857$\pm$37 km/s, given
by Da Costa et al. \cite{dacosta1998},  not in agreement with 894$\pm$29 km/s
 from the
RC3 (de Vaucouleurs et al. \cite{devau1991}), and not in agreement with
 882$\pm$39 km/s from Hilker et al. \cite{hilker1999}. Better agreement is
reached with 836$\pm$32 km/s, the value from the Nearby Early-type Galaxies
Survey (ENEAR, Wegner et al. \cite{wegner2003}). Our value is in very good
agreement with 808$\pm$22 km/s given by Drinkwater et al. \cite{drink2001a}.
As for NGC 1399 itself, 
one finds 9 measurements of its radial velocity  with quoted uncertainties
consulting the NASA/IPAC Extragalactic Database. After skipping two of them (one has a discrepant value and the other a large
uncertainty of 200 km/s), the weighted mean value is 1442$\pm$9 km/s.
The ENEAR-value is 1425$\pm$15 km/s. Our mean radial
velocity of the entire cluster sample is 1441$\pm$15 km/s, so we are confident of
our absolute velocity calibration.

Fig.\ref{distribution} shows the distribution on the sky of all clusters for
which velocities could be measured by cross-correlation. 
 Since not all prepared masks could be observed, some gaps
remain, most strikingly in the South-East quadrant. Due to the
increasing background of the galaxy light, we avoided 
targets with galactocentric distances smaller than about 2$\arcmin$.
Our most distant object lies almost 10$\arcmin$ from the center of NGC 1399.

\subsection{Red and blue clusters}

Because we use 
 the bimodality  in the color distribution of clusters to divide them into two
populations,
we briefly comment on the color-magnitude diagram (CMD). 
A bimodal color distribution is a frequent feature of globular
cluster systems (for recent work see Larsen et al. \cite{larsen2001} and
 Kundu et al. \cite{kundu2001}). In the case of NGC 1399, it has already been 
seen by Kissler-Patig et al. \cite{kissler1997}, Forbes et al. \cite{forbes1998} and Ostrov et al.
 \cite{ostrov1998}.
Fig.\ref{CMD} gives the Washington CMD for almost all of our spectroscopic
sample.
For 32 clusters, we have no photometry for
various reasons, mainly  because these objects were located in 
gaps of the undithered MOSAIC frames.

In this CMD, the color bimodality is not visible. It is, however, very striking
 in the complete
photometry of Paper I.  
But also there, the 
bimodality disappears for objects brighter than R=21 mag and the majority of
these bright clusters are found at intermediate colors.

We  note (as discussed in more detail in Paper I) that the bimodality does not necessarily imply the existence of two and only two
different populations: only the blue (metal-poor) peak lies on the linear part of the
color-metallicity relation and corresponds to a peak in metallicity. On
the red (metal-rich) end, the relation between color and metallicity becomes non-linear and finally flat. This causes a peak at an almost universal color under 
a broad range of metallicity distributions and is not easily interpretable as a
corresponding peak in metallicity. 

Whatever the nature of the bimodality is, it is well known from many studies
that metal-rich clusters behave differently from metal-poor clusters. 
For example, the blue clusters tend to show a flatter spatial distribution than the red clusters (see Harris
\cite{harris2001} for a review). Furthermore, the kinematics tend to be
different between the red and blue clusters.
In the M87 cluster system, for instance,
metal-rich clusters seem to show preferentially radial orbits, while the
orbits of metal-poor clusters are more tangentially biased (C\^ot\'e et al. \cite{cote2001},
but see our corresponding remarks in section 5).
Following Paper I, we  define $C-R$ = 1.6 as the distinguishing color between red and blue clusters. According to the Washington calibration of
Harris \& Harris \cite{harris2002}, this color corresponds to a metallicity of [Fe/H] = -0.6 dex.

\subsection{Velocity distribution of the entire sample}

In Fig. \ref{disptot}, we show the velocity distribution of the entire
sample of 468 objects (upper left panel). The bin size has been chosen to be 70
 km/s, larger than the mean
error, but still of satisfactory resolution. The upper right panel shows an error
 selected
sample ($<$ 50 km/s) demonstrating that the distribution keeps its shape under
error selection. The lower
panel shows an inner sample (left) and an outer sample (right).
Obvious foreground stars with radial velocities of less than 300 km/s are omitted,
 but are listed in the
accompanying data paper.
For the full sample the mean velocity and its  standard deviation  are
1441 km/s and 329 km/s, respectively (note that the velocity dispersions will turn out slightly
 different). This should be compared with the values [1429$\pm$45 km/s; 373 km/s] given by
 Kissler-Patig et al. \cite{kissler1999} from their sample of 76 GCs.

At first glance, the shape of the distribution does not look very  Gaussian. A Kolmogorov-Smirnov test gives
 a probability of only 0.1 that it is drawn from a Gaussian with the above center and dispersion.
The maximum probability of 0.57 is given by a dispersion of 298 km/s.
One non-Gaussian feature is that the distribution does not peak at its mean velocity, a fact
 also seen in the sample of Kissler-Patig et al.  \cite{kissler1999}.
In their Fig. 2, a second peak appears
at a velocity of 1800 km/s. 
Also in our sample a small peak is found at this velocity. It remains there also with lower
bin width, but its significance is doubtful.
The only striking feature is that it is made up of predominantly blue clusters which are
nevertheless evenly distributed over the whole field.
Fig.\ref{disptot} might suggest the existence of two peaks around the systemic velocity, caused by rotation of a subsample. As we will see, no further evidence
for this is found.
 Furthermore,
the distribution looks  asymmetric in that it seems to contain a larger population
of clusters on the low velocity side. This as well is marginally apparent in
Kissler-Patig et al. (their Fig.2), although the authors do not mention it.
To compare the low velocity part (lower than the systemic velocity) with the 
high velocity part, we mirrored both parts to produce two symmetric distributions
and performed Kolmogorov-Smirnov tests. While the high velocities are drawn from a
Gaussian with 1440 km/s as systemic velocity and 308 km/s as dispersion with a
probability of 0.99, the K-S-test finds a probability of 0.63 for the low velocities
using the same parameters. 

However, simple simulations of Gaussian distributions, having the same number of objects, quickly show
that a KS-test frequently gives low probabilities. For the purpose of illustration, Fig.\ref{gaussim}
shows 6 simulations with 470 objects each. The parent populations are Gaussians with 1440 km/s as
the ''systemic'' velocity and a dispersion of 300 km/s. Indicated are the probabilities returned
by a KS-test (we used the command KSTEST/1SAMPLE under MIDAS).  
''Peaks'' occur frequently and one would
conclude that the significance
of peaks in Fig.\ref{disptot} is doubtful as well.     

Fig.\ref{velorad} plots the correlation velocities against projected radii in arcmin. The upper panel shows all clusters,
while the middle panel selects those for which both correlation and line velocities
are available and  differ  
by less than 100 km/s. Again,
it appears as if the low velocity wing would be  more extended,
but also that the limit at  high velocities at 2000 km/s is sharper than at
low velocities. The lower panel shows those objects for which the difference
between correlation and line velocities is less than 50 km/s.
These are our best velocities and here a low velocity limit at about the
symmetric velocity becomes visible as well.
Although one wants to determine the velocity dispersion with as many objects
as possible, the existence of velocity limits symmetric to the systemic velocity suggests that the velocity dispersion should be determined within these limits.
After we have discussed the dynamics,
we will come back to this point and see  that this sharp borders are  about
140 km/s  above
the
circular velocity (415 km/s). Distinctly higher velocities can only be produced by objects on 
very elongated orbits near their pericenters and for which the radial velocity measures 
approximately the space velocity. These are expected to be rare, leaving more
quantitative statements to a theoretical model of the system.  
Possibilities are unbound objects and/or objects in the foreground, which will
be discussed later on. 

 Looking for further
 properties which may distinguish low and high velocities,
we find their  luminosity distribution  quite different.
With respect to velocities,
the cluster candidates have been selected randomly,
and the distribution of magnitudes is accordingly expected to be
similar and symmetric around the systemic velocity. Fig. \ref{velomag}, in
which the R-magnitudes are plotted versus the radial velocities, demonstrates that this is not the case. Indicated by the vertical dashed lines are the
systemic velocity (middle line), a velocity of 2000 km/s
, and
the corresponding symmetric low velocity of 880 km/s. While the number
of objects fainter than R=22 mag statistically agree on the high and
low velocity sides, it is striking that only one object at R=21.6 mag is found
with a
velocity (marginally) higher than 2000 km/s while there are 10 objects with
velocities of 880 km/s or lower with magnitudes brighter than R=22 mag.
This is in spite of the fact that  for both low and high velocities, we are
biased towards bright objects.

The suspicion may arise that some low velocity objects could belong to a
 population
in the foreground, because such asymmetry should
not be present in a spherical and more or less isotropic system. In that case, one would expect these objects not to be concentrated around
NGC 1399. Unfortunately, our sample is not very suitable to search for
differences in the radial distribution of subsamples, since the coordinate
distribution is as much determined by the mask and slit positioning
as by the true spatial distribution. Selecting the objects with, say, less than
900 km/s results in 34 clusters, which must be regarded as the minimum for
senseful statistical statements. We examined 
the radial cumulative distribution of
these clusters together with a comparison sample of velocities higher than
 1300 km/s and found no difference. 

\subsection{Velocity histograms of red and blue clusters}
Fig. \ref{redblue} shows the velocity histograms of the red and the blue
clusters. The total number now is 437 because we do not have
colors for all objects. The distribution of red and blue clusters look rather different.
The formal standard deviations of the velocities are 362 km/s and 289 km/s
 for the blue and red clusters, respectively. As will be  discussed later on, this difference is attributed to
their different surface density profiles. However, these values are larger
than those that we will derive as velocity dispersions, pointing to the 
difficulty in deciding what the correct sample for the determination of
the velocity dispersion is. With this sample, which also includes  the
extreme velocities, we probably overestimate the true velocity dispersion by
the existence of objects which may not belong to the NGC 1399 system. 
The detailed morphology of these histograms is somewhat dependent on the
binning.  However, the peak of
the blue clusters near 1800 km/s is stable against varying the bin width and
it is now clear that it is indeed composed of preferentially blue clusters.
Also the peak at 1550 km/s remains with smaller bin widths than the present 70 km/s.
Whether the central double peak of the red clusters is caused by rotation, will
be investigated later on.

The different velocity dispersions are also visible in Fig.\ref{veloradredblue},
which shows radial distance vs. velocity for the red and the blue sample.
Furthermore, it appears that for distances smaller than about 3 arcmin, the
velocity dispersion decreases towards the center, perhaps somewhat more distinctly
for the red clusters than for the blue clusters. We will come back to this 
point when discussing the velocity dispersions.   

\subsection{Rotation and azimuthal behaviour}

A fundamental kinematic question for our sample is whether we find signatures
of rotation and whether these are different for different subpopulations of globular
clusters. The diagnostic diagram which we have to analyse is a plot of
radial velocities vs. the position angle. C\^ot\'e et al. \cite{cote2001} give a
 useful
discussion of the relation between the intrinsic and projected rotational
velocity field of a spherical system and we do not  repeat that here.
If the intrinsic rotation velocity field is stratified on spheres,
and the galaxy is not seen pole on, we measure radial velocities 
that depend sinusoidally on the azimuth
angle. Therefore, we fit the following relation :
\begin{equation}
  v_r(\Theta)  = v_{sys} + A \sin(\Theta -\Theta_0), 
\end{equation}

where $v_r$ is the projected radial velocity at the azimuth angle $\Theta$,
 $v_{sys}$ is the systemic velocity, and
A the rotation amplitude.

We select velocities between 800 km/s and 2100 km/s to omit the extreme ones.
Fig.\ref{rotrev} plots the radial velocities of three samples vs. the azimuth
angle which goes from North past East. The uppermost panel is the full sample,
then follow the blue and blue clusters. Further selections are the outer
blue clusters (more distant than 6$\arcmin$), the ''double peak'' from Fig.\ref{disptot} for all clusters (between 1280 and 1600 km/s) and the same selection for
the red clusters only, motivated by the right panel in Fig.\ref{redblue} where the
double peak is more prominently visible than for the blue clusters. These samples are
shown in Fig.\ref{rotrot} together with the outer blue clusters.
Least-squares fits for these samples return the values of A and $\Theta_0$
as listed in
Table \ref{rot.tab}.  As the uncertainties (the 1-sigma limits) for the
parameters show, only the blue outer clusters and the ''peak`` selection for
the entire sample show rotation signals. 
Because of the large uncertainties, the rotation amplitude (68$\pm$60) of the outer
 blue
clusters agrees  with the amplitude found by  Kissler-Patig
et al. (1999), but not the position angle (141 $\pm$ 39 degrees East of North).
 These authors   
 found  a rotation amplitude of 153$\pm$93 km/s and $\Theta_0$ equals 
210$\pm$40 degrees (note that the authors give the position angle for the
negative rotation amplitude, not $\Theta_0$). Their  sample comprised  33 clusters more
distant than 5$\arcmin$ from the galaxy center, without distinguishing between
red and blue clusters. According to Caon et al. \cite{caon1994}, the major isophote
axis of NGC 1399 is quite accurately oriented in the East-West direction.
 Within the uncertainties, the outer blue clusters could still rotate
around the major axis.  

However, the fact that the double peak formally shows a rotation signal according to eq.(3)
does not necessarily mean that it is indeed caused by rotation. If this was the case then
one should expect the rotation to be even more pronounced among the red clusters, where the double-peak is more prominent and symmetric with respect to the
systemic velocity.
Fig.\ref{xyplot} selects
 velocities in the intervals 1280 km/s - 1400 km/s and 1480 km/s - 1600
km/s and shows this sample in a x-y-plot (arcmin).  
The horizontal
line indicates the major axis.  Open
hexagons are objects belonging to the lower velocity interval, open triangles
to the higher interval. Additionally, blue clusters are marked by crosses.
This entire sample gives a rotation signal of 20$\pm$8 km/s around the major axis.
However, a large part of the signal comes from the concentration of blue
clusters with low velocities in the South-East quadrant with only a few
high  velocity counterparts in the North-West quadrant. The dominance of
blue clusters on the East side also causes a East-West asymmetry of the
double peak. On the other hand, the red clusters do not show any rotation
signal, but a pronounced double-peak. 
Generally,  rotation does not appear to be significant.

The rotation pattern, resp. its absence, seems to constitute a difference to M87,
 where apparently the entire cluster system  rotates (Kissler-Patig \& Gebhardt 1998).
 The work of C\^ot\'e et al. \cite{cote2001} 
reveals an even  more complicated pattern, where the red clusters rotate around the photometric minor axis, while only the outer blue
clusters rotate around the minor axis and the inner blue clusters around
the major axis. We refer the reader to C\^ot\'e et al. for the interpretation
of this result. 

That only the blue clusters show signs of rotation  was  indicated
also in 
the NGC 4472 cluster system by Zepf et al. \cite{zepf2000}, however, with a low statistical significance. The statistics have been improved by C\^ot\'e et al. \cite{cote2003} who
essentially confirmed Zepf et al.'s result. The metal-poor clusters rotate around the
minor axis, while the metal-rich clusters do not show a significant rotation signal.

\subsection{Measurement of the velocity dispersions}
The most important observable which we want to extract from our data
is the projected velocity dispersion and its dependence on radius.
Given the above considerations regarding the extreme velocities it is clear that
neither measurement nor  interpretation are completely straightforward.

We employ the maximum-likelihood dispersion estimator shown by Pryor \&
Meylan \cite{pryor1993}. We fix  the systemic velocity to 1441 km/s.
Then the velocity dispersion $\sigma$ is calculated (by iteration) according to
\begin{equation}
 \sum \frac{(v_i -v_{sys})^2}{(\sigma^2+ \delta_i^2)^2} =  \sum \frac{1}{\sigma^2+ \delta_i^2},
\end{equation}
 where the sum is taken over all velocities and where the $\delta_i$ denote
 the
uncertainties of the individual velocities. The uncertainty of the resulting 
velocity dispersion is calculated according to the somewhat lengthy expression given
by Pryor \& Meylan.  
We cannot be sure about possible contamination by objects not
belonging to NGC 1399. For example, a couple of objects may be associated with
 nearby dwarf galaxies  and with NGC
1396. Moreover, foreground objects might be present. However, removing 
likely non-members of the NGC 1399 system would require a discussion of individual
objects, which we cannot give here. The global effect on the velocity
 dispersion
will anyway be very small. This, however, is not true for the very extreme velocities
around 600 km/s or 2400 km/s. Fig.\ref{dispvar} shows the dependence of
the velocity dispersion on a lower velocity cut-off (always symmetric to the
high velocity cut-off with respect to the systemic velocity) for the total
sample, the blue, and the red sample. Particularly for the blue clusters,
the effect is not negligible. Since there is no obvious ''correct'' way for
defining, which objects should be considered and which should be discarded,
we refer to the lower panel of Fig.\ref{velorad} and define the upper velocity limit to be
2080 km/s and the corresponding low limit to be 800 km/s. 
However, we also shall show the effect of the full sample on the circular
velocity and the mass profile.

We choose slightly overlapping radial bin widths of 3 arcmin, allowing 
 reasonable statistics to be obtained even
after subdividing the sample according to blue and red clusters.
Table \ref{dispersions} lists the results. 
The correlation velocities
were used to derive  the numbers in Table \ref{dispersions}.
Fig.\ref{dispdat} visualizes Table \ref{dispersions}
and shows the values of the velocity dispersion
$\sigma$ vs. radial distance
for four samples: the unselected sample (uppermost panel), the entire sample
under velocity selection (second panel), and further selections of blue
clusters (third panel) and red clusters (lower panel).
The radial bins overlap to force some smoothing, so the values are not independent. 
The only bin which marginally deviates from a radially constant velocity dispersion
is the innermost bin of the blue clusters.
{\it We therefore consider the projected velocity dispersion to be constant within the uncertainties}. 
Within 3 arcmin radius, there is some indication that the dispersion decreases
towards the center. Using the same velocity selection, the dispersion for
radial bins of $<$2$\arcmin$, 2-2.5$\arcmin$, 2.5-3$\arcmin$ are
219$\pm$29 km/s, 211$\pm$24 km/s, 292$\pm$31km/s, respectively. Using the
entire sample, the corresponding values are 263$\pm$34 km/s, 318$\pm$34 km/s,
331$\pm$35 km/s. However, the sample sizes are small and whether this
decrease is real or not, needs further confirmation.

The results stand in some contrast to the results of Kissler-Patig et al. \cite{kissler1999},
 who suggest a strong increase between 2 and 10
arcmin. Their radial bin samples are very small and have large 
uncertainties, which individually overlap with our results. Ignoring their
innermost bin at 2 arcmin,  their work also supports a  
constant dispersion,  but at a value of 373$\pm35$ km/s which we would not
reach even without any selection. 
All differences are most likely due to their small numbers, large measurement uncertainties and also 
inhomogeneous data. In Paper III we show that the various sources of the velocities used by  Kissler-Patig et al.
have considerable systematic velocity shifts. Therefore, we do not discuss these differences further.

Another point in Fig.\ref{dispdat} is striking: the red clusters exhibit a
systematically lower dispersion than the blue clusters. We shall see in more detail in
the following section on dynamics that this can be explained by the difference in their
spatial density profiles.

\section{Dynamics}

Our intention in obtaining this large velocity sample was to have a sufficient
number of probes to derive the mass profile of NGC 1399 and the orbital structure
simultaneously without having  to make restrictive assumptions concerning
spherical symmetry or isotropy.
The most general models come from orbit-based axisymmetric analysis (Gebhardt et al. \cite{gebhardt2000}).
However, these models require a significant parameter space to explore which
we will consider in a subsequent paper (Gebhardt et al. \cite{gebhardt2004}).
In this paper, we will first consider spherical models with different anisotropies,
 and discuss
possible biases that this may create.

There are good reasons to believe that NGC 1399 is well approximated by
a spherical and more or less isotropic model. Moreover, the fact that the projected velocity dispersion does not
change with radial distance indicates that these properties as well do not change
significantly with radial distance. A spherical model was also used by
 Saglia et al. 
\cite{saglia2000} and Kronawitter et
al. \cite{krona2000} which allows us to make a direct comparison.

\subsection{Spherical models}

To provide the nomenclature, we briefly introduce the spherical Jeans equation. Binney \&
Tremaine \cite{binney1987} showed how it can be derived from the collisionless Boltzmann-equation.
It reads:

\begin{equation}
 v_{circ}^2 =\frac{G \cdot M(r)}{r} = - \sigma_{r}^2 \cdot (\frac{\mathrm{d}\ln\rho}{\mathrm{d}\ln r}
+ \frac{\mathrm{d}\ln\sigma_{r}^2}{\mathrm{d}\ln r} + 2 \beta), 
\end{equation}

where $v_{circ}$ is the circular velocity, $G$ is the constant of gravitation, $r$ the galactocentric distance, $M(r)$ the
mass contained within $r$, $\sigma_r$ the radial component of the velocity dispersion, $\rho(r)$
the density profile of clusters, and $\beta = 1- \frac{\sigma_{\Theta}^2}{\sigma_{r}^2}$ with
$\sigma_{\Theta}$ being the tangential velocity dispersion. The azimuthal velocity
dispersion $\sigma_{\Phi}$ is equal to $\sigma_{\Theta}$ in the spherical case.

Not all parameters are accessible from our observations, which only measure {\it projected } values. While we can determine 
$\frac{d\ln\rho}{d\ln r}$ from our wide-field photometry of the GCS (Paper I),
we have no straightforward possibility to deproject $\sigma_p$, our observed projected velocity
dispersion, because we only observed clusters out to 9 arcmin, while the wide-field photometry
still finds cluster candidates at 20 arcmin distance.

We  assume (as all previous workers did) that NGC 1399 exhibits spherical
symmetry. Strictly, this is not true.  The ellipticity of NGC 1399 ($\epsilon
= 1-b/a$) is modest, but non-zero. It ranges from about 0.1 in the inner region
and may increase to 0.2 (Caon et al. \cite{caon1994}, Paper I). 

If NGC 1399 deviates from our spherical assumption, the likely effect would be for us
to underestimate the mass 
(an oblate elliptical, supported by
an anisotropic velocity dispersion, exhibits a smaller velocity dispersion than
would correspond to its mass in a spherical configuration). As Magorrian \& Ballantyne \cite{magorrian2001} show, this situation
is also expected to create a spurious radial bias. In the sample of 21 round elliptical
galaxies of Gerhard et al. \cite{gerhard2001}, which they analyzed under the assumption of spherical
symmetry, NGC 1399 has a central $M/L_B$ value of 10, one of the highest occuring. It
thus seems unlikely that this value  for NGC 1399 is underestimated. However, 
Gerhard et al. state that no correlation is seen between their central $M/L$-values
and the anisotropy. A $M/L_B$-value of 10 may not be surprisingly high, but so
 far, old metal-rich populations like the galactic globular clusters 
NGC 6496 and NGC 6352 do not show the steep mass functions which are needed
to produce such a value. Evaporation of low mass stars also is no likely
explanation (Pulone et al. \cite{pulone2003}). 

We have seen that the red and the blue clusters have different velocity
 dispersions. They
must trace the same mass. If we
set $\frac{d\ln\sigma_{r}}{d\ln r} = 0$, i.e. adopting a radially constant $\sigma_{r}$
we have 

\begin{equation}
 \sigma_{r,red}^2 \cdot \big(\frac{\mathrm{d}\ln\rho_{red}}{\mathrm{d}\ln r} + 2\beta_{red}\big)  =
 \sigma_{r,blue}^2 \cdot \big(\frac{\mathrm{d}\ln\rho_{blue}}{\mathrm{d}\ln r} + 2\beta_{blue}\big)
\end{equation}

The (deprojected) density slopes are $-2.64\pm0.1$ and $-1.8\pm0.1$ for the red and the blue
clusters, respectively (Paper I). For the red clusters, these
numbers are valid for a larger radial range than we are considering here, while the 
blue clusters assume the slope of the red clusters for r$>$8$\arcmin$. We see from
the above equation that if $\beta$ would be zero for both populations, the difference in the velocity
dispersion accounts for the different radial density profiles: The equation predicts
$\sigma_{r,red}/\sigma_{r,blue} = 0.83$, while the observed value is 0.88.

We cannot deproject the velocity dispersions with our still noisy data and our incomplete coverage
of the cluster system. As Paper I shows, the cluster system extends to at least 100 kpc
in radius (recall that the most distant Milky Way globular clusters are beyond 100 kpc). But we could get a feeling about  projection effects
 by projecting a model cluster system with different anisotropies and
requiring  it to
give a constant projected velocity dispersion.
Doing so, we assume a cluster system with red and blue clusters, which has density profiles with
power law exponents -2.64 and -1.8, respectively. 

The following projection formula is used:

\begin{equation}
 \sigma_p^2(R) = \frac{2}{N(R)}\cdot \int_R^{R_0} n(r) \sigma_r^2(r) (1-\beta(r) \frac{R^2}{r^2})
 \frac{r dr}{\sqrt{r^2-R^2}}, 
\end{equation}

where $\sigma_p$ and $\sigma_r$ are the projected and radial velocity dispersions, respectively,
and $N(R)$, $n(r)$ the density profiles in the same sense.
Now we adopt $n(r) \sim (1+r/r_c)^\alpha$, where $r_c$ is a core radius, small compared with the inner
limit of 10 kpc, so that a pure power-law density profile is achieved already
at this radius. $\alpha$ is an exponent which takes the above values for red and blue clusters,
respectively. We also adopt an outer cut-off radius, $R_0$, which is set to 100 kpc, corresponding to 18 arcmin.
This choice is arbitrary and assumes that clusters outside this radius, if they
exist, do not significantly influence the projection effects.
We then assume for the two values of $\alpha$ different, but radially constant values for the anisotropy in order to avoid a further extension of  parameter space.  
$\sigma_r$ is assumed to vary linearly between 10 kpc and the cut-off radius,
i.e. $\sigma_r = a_1 \cdot r + a_2 $. For
the isotropic case, $\sigma_r$ is radially constant.
 
Now we ask: for each value of the anisotropy,  what
radial dependence is required for  $\sigma_r$
to produce  {\it constant projected velocity dispersions of 291 km/s for the blue clusters and 
255 km/s for the red clusters}?
After having found (by trial and error) those values of $a_1$ and $a_2$ which
in projection give constant velocity dispersions within a few km/s, 
 we then can calculate the circular velocities 
according to the Jeans equation. 

The results of this calculation are shown in Fig.\ref{dispproj}. 
The upper panel refers to the red clusters, the lower panel to the blue clusters. 
 The
different values for $\beta$ are indicated  
at the right hand side ($\beta$ = 0.8 has been skipped for the blue clusters). We overplot the circular velocity resulting
from the luminous component alone. 
Table \ref{coeffs} lists the corresponding values of $a_1$ and $a_2$. If we
assume isotropy for the red clusters, we get a constant circular velocity of
415 km/s. 

To construct a mass model for the inner region of NGC 1399, we need
the (deprojected) luminosity density profile and the assignment of
a $M/L$-value. Neither of which can be achieved in a straightforward manner.

For the luminosity density profile we adopt the R-profile given in
Paper I. 
 Since the surface brightness profile itself
is a composition of different observations done in different photometric
bands, one has to adopt shifts depending on the color of NGC 1399 (see Paper
I for details).
Our adopted R luminosity density profile (Kron-Cousins) is equation (2),
which corresponds to a distance of 19 Mpc.

Saglia et al. \cite{saglia2000} quote from their dynamical modelling a $M/L_B$-value of 10. To convert this into a $M/L_R$-value, one has to know the B-R colors
of NGC 1399 and the sun (Cousins-system). B-R colors (in various apertures) are given by
 Poulain \cite{poulain1988} (1.56), Sandage
 \& Visvanathan \cite{sandage1978} (1.47), Mackie et al. \cite{mackie1990} (1.4),
 Lauberts \cite{lauberts1984} (1.62). We adopt a value of 1.55, omitting 
Mackie et al.'s measurement.
Regarding the sun we use the theoretical B-R color for the appropriate effective
temperature of solar-like stars of Houdashelt et al. \cite{houda2000} and adopt
$\rm (B-R)_{sun} = 0.97$. This gives a factor of 0.59, by which $M/L_R$ is smaller
than $M/L_B$. In the following, we use $M/L_R$=5.5, taking into account the
different adopted distances. 

Also overplotted are the total circular velocities resulting from the sum of the luminous potential
and the best fitting dark halo model (which we describe below).

The main effect of varying $\beta$ for the clusters is that in the case of a radial anisotropy ($\beta > 0$), little mass is needed to produce the required constant projected velocity dispersion, while more mass is needed when the
anisotropy changes to tangential ($\beta < 0$).
That this effect is larger for the blue clusters than for the red clusters is a
 result of their shallower density  profile (the density profile and $\beta$ are
 additive in the Jeans equation). We note that in reality the effect of $\beta$
on the blue clusters might be slightly less than indicated in the figure. The
reason is that the density profile of the blue clusters becomes steeper and
undistinguishable from that of the red clusters (Paper I) beyond 50 kpc of
projected radial distance, while for this
calculation we extrapolated the shallower profile of the inner region out to the
cut-off radius.  

The uncertainties of the derived circular velocities are difficult to formalize.
Apart from the dominant systematic effect of the application of velocity limits,
they depend on the distance, on the uncertainty of the surface density profiles,
on the assumption that $\beta$ really is constant for the entire cluster system
and varies in the adopted manner. It is easier to follow the propagation of the
observational uncertainty, represented by the uncertainty of the projected
velocity dispersion. To give a number, an uncertainty of 15 km/s translates
roughly into 30 km/s in the circular velocity. It is then interesting to note
that for the red clusters the degeneracy between mass profile and orbital 
structure is not more important for the derivation of the mass profile
than are the observational uncertainties.  In other words: a population with
a steep density profile is sensitive to mass, a population with a shallow 
density profile is sensitive to the orbit structure. 

The above mentioned relation between the $\beta$-values for the two populations is also
 visible in  Fig.\ref{dispproj}. For isotropy, the blue clusters give,
within the uncertainties, the same circular velocity as do the red clusters.
The slightly stronger tangential behaviour of the blue clusters is not really
significant. If the blue clusters were more tangential then also the circular
velocities for $\beta = 0$ would be displaced to higher values. So the
different circular velocities for assumed isotropy  measure 
the difference of the $\beta$-values. We therefore can say with some confidence
that both populations are close to isotropic. 

How does the analysis of the globular cluster system compare with the analysis
of the inner stellar body of NGC 1399? 
  As did Saglia et al. \cite{saglia2000},  we use
for the dark halo a  logarithmic potential of the form 

\begin{equation}
 \Phi_{DM} = \frac{1}{2}\,  v_0^2\, ln(r^2+r_0^2), 
\end{equation}
where $v_0$ is the asymptotic circular velocity and $r_0$ the halo core radius.

Because of $ v_{circ}^2 = r\, d\Phi/dr $ , we have
\begin{equation}
 v_{circ}^2 = \frac{v_0^2}{1+(\frac{r_0}{r})^2} \hspace{0.5cm}. 
\end{equation}

To represent the dark halo by a logarithmic potential, we choose the circular
velocities derived from the red clusters because of their lower sensitivity to  $\beta$ and
assume isotropy. Then we subtract the  luminous component
 from the total mass and get the circular velocity curve for the dark halo
 alone
at the four radii between 10 kpc and 40 kpc given by Fig.\ref{dispdat}.
A least-square fit is used to obtain $v_0$ and $r_0$. 
We get $v_0 = 365\pm6$ km/s and $r_0 = 11.7\pm0.7$ kpc. 
The fit uncertainties are small and do not reflect the true systematic
 uncertainties, 
 which are difficult to parametrize. But they do account for the
observational uncertainty of the projected velocity dispersion. 
Systematic uncertainties lie in the adoption of M/L and the determination
of the circular velocity. 
Small variations of M/L ($\Delta M/L_R < 1$) do not change $v_0$ 
within the above uncertainties, but have a stronger effect on  $r_0$:
one gets $\Delta v_0/\Delta (M/L_R) = 3$. Variations of the 
circular velocity change $v_0$ approximately by the same amount and let
$r_0$ vary as $\Delta v_0/ \Delta v_{circ} = -0.06$.

The ``best fitting model'' of Saglia et al. \cite{saglia2000} has $r_0 = 210 \arcsec$ (19.3 kpc) and $v_0 = 323$ km/s as the dark halo parameters. For other models, which also fit well the Saglia et al. data
and even better to our circular velocities, no dark halo parameters are quoted.

Kronawitter et al. \cite{krona2000} use the same data as Saglia et al., but different models and a different
distance (21.9 Mpc). They do not list their model parameters, but give the range of acceptable 
circular velocities at the last data point. These models seem to fit well, although the range still is
appreciable. We note that the ''best'' $M/L_B$ of Kronawitter et al. is 10.6,
i.e. larger than that of Saglia et al.'s in spite of the larger distance which
would one let expect a proportionally smaller $M/L_B$.
 
If we  use this interval as reference and find those halo parameters of logarithmic
potentials which span the acceptable range, we get the following: 

\noindent
Model 1 (high): $v_0$ = 390 km/s, $r_0$ = 7.7 kpc\\
Model 2 (best): $v_0$ = 370 km/s, $r_0$ = 10.5 kpc\\
Model 3 (low) : $v_0$ = 350 km/s, $r_0$ = 19.2 kpc\\

These dark matter models, to which we added our luminous component ($M/L_R = 5.5$), are plotted in Fig.\ref{dispproj2} (dashed lines) together with our
best model (solid line) and that of Saglia et al. (dotted line). The latter one is slightly radial.
The Saglia et al.-model which is almost isotropic within our radial range, has a circular velocity
of 404 km/s at 8.9 kpc. 
It is encouraging to see how well the results agree in spite of the completely
different analyses.  

Had we renounced any velocity selection, the agreement would be worse.
The unselected samples of blue and red clusters have velocity dispersions of
350$pm$14 km/s and 287$pm$15 km/s, respectively. These values
result (for isotropy) in a constant circular velocity of 465 km/s. A fit
of a logarithmic potential gives $v_0$=412 km/s and $r_0$=9 kpc. The
dashed-dotted line in Fig.\ref{dispproj2} shows this model. There is no
hard evidence to reject it, but the difference to the stellar models is
considerable. Moreover, the increase of the circular velocity between 5 kpc
and 20 kpc is more pronounced than in the other models. In order to have 
a constant projected velocity dispersion, the anisotropies of the blue and
red clusters must therefore conspire.
But again, the present data does not allow a firm exclusion.

\section{Discussion}

\subsection{Comparison with earlier work on NGC 1399}

As mentioned before, there have been already some attempts to
investigate the kinematics of the NGC 1399 GCS. The most recent work is that
of Kissler-Patig et al. \cite{kissler1999}, who compiled the radial velocities
 of 74 GCs. A comparison of their velocities in common with ours 
can be found in Paper III. The main finding of Kissler-Patig et al. is that
the dispersion increases from about 260 km/s at  2$\arcmin$ to almost 400 km/s at 
 8$\arcmin$ of radial distance. The similarity of the outer velocity dispersion with that of Fornax galaxies leads them to conclude that the outer clusters
might be moving in the overall Fornax potential rather than in the potential
of NGC 1399. However, the uncertainties in the velocity dispersion, caused by the low number of objects and relatively large velocity uncertainties,
 are so large that even a constant velocity dispersion of the order 300 km/s is covered by almost all
of their error bars. Kissler-Patig et al. note a marked turn-up in the dispersion
 between the 
stellar component and the clusters. We rather say that this depends on what cluster population is considered and also what source
for the stellar velocities is used. Kissler-Patig et al. use older data by Franx et al. \cite{franx1989},
Bicknell et al. \cite{bicknell1989}, and Winsall \& Freeman \cite{winsall1993}. The more recent study by
Saglia et al. \cite{saglia2000} finds about 250 km/s as the average of their outermost point which is in excellent agreement with
the red clusters, as one would expect, because this population best represents 
the stellar component both in terms of color and radial profile, whose power-law exponent
is -1.8 (see Paper I).

The statement that the clusters are moving in the general Fornax potential can have a dubious meaning.
 The equality
of the  velocity dispersions of GCs and galaxies, however, does not say much about the dynamical relationship of
these components. For example, Drinkwater et al. \cite{drink2001} give the quite different values of $308\pm30$ km/s and $429\pm41$ km/s
as the velocity dispersions for giant and dwarf Fornax galaxies, respectively (they interpret this as the
difference between virialization of the giants and infall of the dwarfs).   

 Of course, we assume that the majority of the clusters we observed are bound to NGC 1399.
 But we do not know how far the dark 
halo extends and we cannot exclude that NGC 1399 is located at the bottom of a ``cD dark halo'' which is
the central substructure in a cluster-wide dark halo (Ghinga et al. \cite{ghigna2000}). In this sense,
the halo of NGC 1399 would be part of the cluster potential, but finding evidence for this is beyond the
present observational possibilities.

Turning to planetary nebulae, Napolitano et al. \cite{napo2002} re-analyzed older velocity data by Arnaboldi et al.
\cite{arna1994}. In their sample of 37 PNe, they see evidence for a disturbed velocity structure and
assign this to a possible past encounter with NGC 1404. We cannot comment 
 on that apart from saying that we do not see it, but it would be
very interesting to have a larger sample of PNe to prove or disprove it.

\subsection{The mass profile}
\subsubsection{General remarks}
NGC 1399 has been sometimes labelled as the central cD galaxy in the Fornax
cluster, expressing the fact that its surface luminosity falls off slower
than a de Vaucouleurs law.
The derived total  mass profile resembles that of an isothermal sphere: its
total density falls off like $r^{-2}$. Its luminous surface density in the R band
 falls off like $r^{-1.85}$ (Paper I) which is consistent with a deprojected mass profile describable by a uniform power law with an exponent of
$r^{-3.0}$, if the color gradient is taken into account. No feature, which
would indicate the onset of a ''cD-halo'' is discernible. 
 As Jaffe \cite{jaffe1987} and White \cite{white1987} already have 
argued, a $r^{-4}$ profile in the outer regions of elliptical galaxies emerges,
 if stellar orbits are scattered near the escape
energy in the presence of a nearly Keplerian potential. In projection, this
behaviour well resembles a de Vaucouleurs profile.
In the case of NGC 1399, we have an isothermal sphere  out to about 4 effective radii, and since the $r^{-3}$ profile of NGC 1399 is realized out to at least
10 effective radii, one may conjecture that also the isothermal sphere extends
to this galactocentric distance.
Under this view, NGC 1399  would be distinguished from ``normal elliptical
galaxies'' with a de Vaucouleurs profile by the fact that due to the huge dark halo, the Keplerian regime
sets in at much larger distances than we are able to investigate and most
stars have energies much lower than the escape energy.
We note that only a $r^{-3}$ profile brings in agreement the velocity dispersions
derived from the Jeans equation and from the virial theorem (see van den Bosch 
\cite{bosch1999} for remarks on that matter) in case of an isotropic tracer population in an isothermal sphere. It would be very interesting to also investigate the nearby central galaxies like M87 and NGC 3311, but accurate wide-field
photometry is not yet available for them.

\subsubsection{The dark matter profile}

The question arises whether the dark halo of NGC 1399 could be identified
with a dark halo predicted from Cold Dark Matter (CDM) simulations. Such
simulations make specific predictions regarding the structural properties,
especially the density profile, of dark halos. However, in the context of
galaxy formation one expects a CDM halo not to maintain its structure, but
to be modified according to the dynamical details of the formation history. 
Therefore our intention is not to prove or disprove CDM predictions, but to
morphologically describe the dark halo in terms of parameters used in CDM
simulations. 

We adopt the  density profile as 

\begin{equation}
\rho(r) \sim  \frac{\rho_s}{(r/r_s)^{\zeta} (1+(r/r_s))^{3-\zeta}}, 
\end{equation}
where $\rho$ is a characteristic density and $r_s$ a characteristic radius. 
The case $\zeta = 1$ describes halos of the ''NFW''-type (Navarro et al. 
\cite{navarro1996}, \cite{navarro1997}). At small radii, its density profile
is proportional to $r^{-1}$, at large radii to $r^{-3}$.
Other workers like
Ghigna et al. \cite{ghigna2000} and Klypin et al. \cite{klypin2001},  
performing N-body simulations of higher resolution, agree well regarding the outer regions, but predict steeper profiles 
($\rho(r) \sim r^{-1.5}$) near the center.

For the cases $\zeta=0$ and $\zeta=3/2$, the mass $M(r)$ is as well analytically
integrable and we will also consider them ($\zeta=2$ does not result in a viable
 description). 

In general, M(r) becomes
\begin{equation}
M(r) = 4\pi \rho_s r_s^3 g(x), 
\end{equation} 
where $x = r/r_s$ and $g(x)$ is for the different values of $\zeta$:

\begin{eqnarray}
g(x) = & \ln(1+x) + \frac{2}{1+x} - \frac{1}{2(1+x)^2} - 3/2& (\zeta=0)\\
g(x) = & \ln(1+x) - \frac{x}{1+x}                           & (\zeta=1)\\
g(x) = & \frac{2}{3} \ln(1+x^{\frac{3}{2}})             & (\zeta=\frac{3}{2})\\
\end{eqnarray} 

In terms of circular velocities and convenient units:
\begin{equation}
v_{circ} [\mathrm{km/s}] = 2.1\cdot 10^{-3} \sqrt{\frac{M(r)}{r}\frac{\mathrm{kpc}}{M_{\odot}}},
\end{equation}

Adopting isotropy for the red globular cluster population (the blue cluster
population then becomes slightly tangential) results in a circular velocity
of 415 km/s between 10 and 40 kpc of galactocentric radius. From this we
subtract the luminous component and make least square fits to the various
 profiles
as we did for the logarithmic potential.

Within our radial range, the cases $\zeta$=0 and $\zeta$=1
fit equally  well to the circular velocity
curve of the dark matter. $\zeta$=3/2 fits badly and we do not consider it further. For $\zeta$=0, $r_s$ is a core radius and $\rho_s$
the central density. We find $r_s$ = 11.6$\pm0.7$ kpc and $\rho_s$ = 0.14$\pm$0.001 $M_{\odot}/pc^3$.
For $\zeta$=1 (NFW-profile), the fit results in $r_s$ = 33.5$\pm$1 kpc and
 $\rho_s$= 0.01$\pm$0.0006 $M_{\odot}/pc^3$. 
The formal fit errors only say how well the points fit to the adopted
density profile, but 
the important uncertainties come from the uncertainties of both the circular 
velocities and the adopted M/L of the stellar population, which we here
assume to
be constant, but may well decline with radius indicated by the color gradient
in NGC 1399 (Paper I).     

Let us assume that we are dealing with an NFW-halo. Navarro et al.
\cite{navarro1996}, \cite{navarro1997} noted that the density and the scale parameter
are strongly inversely correlated. The reason is that  small halos were formed earlier 
when the universe was denser than at later times. This led to the concept of the
universality of CDM halos, the halos being described by only one 
parameter, for instance its mass. However, other simulations (Bullock et al. \cite{bullock2001}, Jing \cite{jing2000},
 Jing \& Suto \cite{jingsuto2000})
questioned the universality or at least showed a much larger diversity.

Following Bullock et al., more customary parameters than $\rho_s$ and $r_s$ are the virial mass $M_{vir}$,
which is enclosed by a  virial radius $R_{vir}$, inside which the
mean density is by a certain factor higher than the mean universal density,
depending on the cosmological model. The related velocity is $V_{vir}^2 =
G M_{vir}/R_{vir}$. For the cosmological parameters $h$=0.7,  $\Omega_m$=0.3,
and an overdensity factor of 337 (see Bullock et al. for details),
handy relations are $M_{vir} = 0.64\cdot 10^5 R_{vir}^3$
and $V_{vir} = 0.53\cdot R_{vir}$ with $M_{vir}$ in solar masses,
$R_{vir}$ in kpc and $V_{vir}$ in km/s.

Besides $M_{vir}$, the concentration $c_{vir} = R_{vir}/r_s$ is the second
parameter. Then one has in terms of circular velocities of the dark halo:

\begin{equation}
V_{circ}(r)^2 = V_{vir}^2 \cdot \frac{c_{vir}}{x}\,\frac{g(x)}{g(c_{vir})},
\end{equation}
where $x=r/r_s$ and $g(x)$ is the above relation for the NFW-profile (equation 13).

We find that a dark halo with $M_{vir} = 9.7 \cdot 10^{12} M_{\odot}$, $R_{vir}
= 533$ kpc,
and $c=15$  is
the best representation of our circular velocity curve. Rather than 
giving the formal fit uncertainties (which are of the order 2\% in $R_{vir}$
and 5\% in $c$),
we vary M/L and the constant circular velocity of the total mass to
provide estimations, how these parameters influence the halo properties. 
For small changes of $M/L_R$ up to 1,
we find $\Delta M_{vir}/\Delta \frac{M}{L_R} = 2.8\cdot 10^{12}$ and
 $\Delta c/\Delta \frac{M}{L_R} = -4.5$. Variing $V_{circ}$ between 390 km/s
and 450 km/s has a
small non-linear effect on $ M_{vir}$ ($9.1 \cdot 10^{12} < M_{vir} < 1.1 \cdot 10^{13}$), and 
$\Delta c/\Delta V_{circ} = 0.15$. For example, if $V_{circ}$ would be 440 km/s
instead of 415 km/s, the concentration parameter would change to 18 instead
of 15. 
These relations mean that an uncertainty of, say, 30 km/s in the circular 
velocity  affects both $M_{vir}$ and $c$ only modestly.
The same holds true for $M/L_R$. 

Bullock et al. \cite{bullock2001} show in their Fig.4 the relation between
the concentration parameter and $M_{vir}$ for their simulations. This plot
reveals (for the Hubble parameter h=0.7) that a dark halo with a mass
of $M_{vir} = 9.7 \cdot 10^{12} M_{\odot}$ and the concentration $c=15$ 
falls more or less on the mean
relation between $c$ and $M_{vir}$.  It is still consistent with being
a CDM halo. It is, however, obvious that galaxies like NGC 1399 with a high
luminous matter density in the inner region are not very suitable for proving
or disproving, whether an observed dark halo resembles the type of halos
found in CDM simulations.

 Fig.\ref{CDM} shows
the various circular velocity curves for these parameters. The dotted
line is the luminous component, the solid lines represent the circular velocities 
for the NFW halo alone and its sum with the stellar profile, respectively. The dashed-dotted lines describe the same for the logarithmic potential. They are practically
identical except for the inner cuspy behaviour of the NFW profile.
 For comparison, the circular velocity
curve of Kronawitter et al. \cite{krona2000} (their ''best fit``, read off from
their Fig.18) is indicated by crosses.  
However, given the various sources of uncertainties, starting from
the absolute photometric calibration to the transformation to $M/L_R$, one
only can state that the agreement is quite good. The upper 95\% confidence limit
of the Kronawitter et al.-profile encompasses both the NFW halo and the logarithmic
potential  so one cannot distinguish between these two.

On cluster scales dark matter halos seem to be quite consistent with NFW
halos (McLaughlin \cite{mac1999}, van der Marel et al. \cite{vandermarel2000}).
However, recent observations of brightest cluster galaxies (Kelson et al.
\cite{kelson2002}, Sand et al. \cite{sand2002}) resulted in much shallower
dark matter profiles which apparently challenge the universality of NFW profiles. Also in low surface brightness galaxies, de Blok \& Bosma \cite{deblok2002} find cores rather than the predicted cusps.   

\subsubsection{The comparison with X-ray mass profiles}
Fig.\ref{massprof} compares the various X-ray mass profiles of Jones et al.
 \cite{jones1997} (ROSAT,PSPC), Ikebe et al. \cite{ikebe1996} (ASCA), and
Paolillo et al. \cite{pao2002} (ROSAT,HRI) with our logarithmic
potential out to 50 kpc of galactocentric radius. The Jones et al.  profile
(long dashed line) has been obtained using their quoted formula with a 
constant temperature of 1.3 keV and with a value of 1.27 for the exponent
of the gas density distribution, corrected for a distance of 19 Mpc. The 
ASCA profile (dot-dashed line) has been read off  from Ikebe et al.'s diagram, also assuming
a linear dependence of the cumulative mass on the distance. Paolillo et al.'s
profile is the short dashed line, read off from their Fig.17 (power-law
temperature profile).

 The solid line
is our mass profile for $M/L_R$ = 5.5 and $V_{circ}$ = 415 km/s. Two dotted lines indicate mass profiles, resulting from
the assumption of circular velocities of 450 km/s and 390 km/s, respectively.
These values might be considered as the tolerance range. The lower dotted
line is the luminous component. Since it has $M/L_B = 10$, one reads off
$M/L_B = 46$ for 50 kpc. While the ROSAT
profile is in quite good agreement with our mass profile, the ASCA profile is
characterized by a 
flattening which beyond 80 kpc turns up again, suggesting to Ikebe et al.
the existence of substructure in the dark matter distribution of the Fornax
cluster. 

Substructure  is also present in the X-ray mass profile of Paolillo et al.
\cite{pao2002} (short-dashed line). These authors describe the X-ray halo by
three components which they label a core, a galactic, and a cluster component. Within
the radial range $2\arcmin < r < 8\arcmin$, the galactic component dominates
with a mass profile approximately like $M(r) \sim r^2$ (their Fig.17). This is not in agreement
with our constant velocity dispersion, which implies $M(r) \sim r$. In the
isotropic case, a quadratic mass dependence on radius would require
$\sigma \sim r^{1/2}$, i.e. $\sigma$ should increase by a factor of 2, which
is ruled out. The Ikebe et al. profile predicts just the opposite, namely a
decrease of the velocity dispersion, which again we do not see. We can only
suspect without having hard evidence that the non-spherical and irregular shape
of the X-ray halo might be the reason for the bending of these mass profiles.
It is also noteworthy that no corresponding feature is visible in the stellar
luminous profile. 
The investigation of more distant clusters certainly would be helpful.

A comparison of Fig.\ref{massprof} and Fig.\ref{dispproj} leads to the
conclusion that the global orbit distribution of the clusters should be close to
 isotropic as has been found also in the cases of M87 (C\^ot\'e et al.\cite{cote2001})
 and M49 (C\^ot\'e et al.\cite{cote2003}). We
cannot really constrain the anisotropy of the red clusters, but the blue clusters
trace the same mass, which narrows down the permitted range of anisotropies.
Even without a detailed analysis of the higher orders of the velocity distribution,
a look at Fig.\ref{redblue} suggests that the blue clusters are more
 tangential than the red clusters, certainly not the opposite. A true circular velocity of, say, 450 km/s instead
of 415 km/s would fit even better at least to the ROSAT X-ray mass, but according to
Fig.\ref{dispproj} would cause the
red clusters to gain a higher tangential bias than the blue clusters. On the other
hand, lowering the circular velocity would leave the blue clusters more tangential,
but enhances the difference to the X-ray mass. 

\subsubsection{Extrapolation of the mass profile}

It is interesting to extrapolate the mass profiles of the NFW halo and
the logarithmic potential towards larger galactocentric distances in order
to see how they would fit into the cluster mass profile. Fig.\ref{masslarge} shows the mass profiles out to 500 kpc. The numbers 1,2,3 label the 
ROSAT mass profile, the mass profile of the Fornax cluster derived from galaxy velocities (Drinkwater et al. \cite{drink2001}), and the ASCA mass profile, respectively. 
Labels 4 and 5 mark the logarithmic potential and the NFW halo.
The latter two halos are hardly distinguishable out to 200 kpc. A comparison
with the  Fornax cluster mass profile 
reveals that neither the
logarithmic potential nor the NFW halo
can be the dominant component of this mass, which at 500 kpc is (3$\pm$1$)\cdot10^{13} M_{\odot}$.
More than 40 galaxies, roughly half of Drinkwater et al.'s sample, are projected
 within this radius. According to Drinkwater et al., this corresponds to 
$10^{11} \mathrm{L_{\odot}}$ (B-band). Even to achieve only marginal
agreement one would need for these galaxies an average $M/L_B$ of 100 in case
of the NFW halo and 50 in case of the logarithmic potential. These are lower
limits considering that the
true number of galaxies within 500 kpc is less than the projected number. The ASCA and ROSAT profiles (disregarding their mutual disagreement in shape) 
agree with Drinkwater et al.'s analysis. Any NFW halo which could
account for the mass of this analysis, is not consistent with the halo
near NGC 1399.

Several reasons for this discrepancy are thinkable: the dark halo of NGC 1399
is not of the NFW type or it is only a substructure within the global Fornax potential, 
or the mass profiles are still not correct. The radial range 50\,kpc$<$r$<$100\,kpc seems to be sensitive 
for disciminating between the two X-ray profiles. Again, the analysis of globular cluster
velocities at larger radii will be of high interest.

\subsection{Do all objects belong to NGC 1399?}

As already mentioned, one can suspect a few clusters  
to be members of dwarf galaxies close to NGC 1399. Interestingly,
two of the objects with velocities higher than  2000 km/s seem to belong to
FCCB 1241 (Ferguson \cite{ferg1989}), which Ferguson lists as a possible member
of the Fornax cluster.  But also general considerations regarding the extreme
velocities leads one to suspect that not all objects are members of the NGC 1399
system.

Radial velocities below 800 km/s or higher than 2080 km/s, which we observe
out to 40 kpc galactocentric radii, mean velocities higher than 640 km/s
relative to NGC 1399. These velocities must belong to objects  near their
perigalactocentric distances. Let us assume that they are exactly at their
pericenters and that we  observe their space velocities (which by itself
seems to be very unlikely). Then with the knowledge of the potential we
can ask what minimal apogalactocentric distances these objects must have. Doing so
we approximate the sum of luminous and dark matter by a logarithmic potential
and find that the parameters $v_0 = 400$ km/s and $r_0 = 0.7$ kpc are
a reasonable representation. Moreover, we assume that this potential can be
extrapolated to large radii out to, say, 400 kpc.

Considering only bound orbits, conservation of energy and angular momentum
gives the apocentric and pericentric velocities and radii $v_a$,$v_p$,$r_a$,$r_p$.
If $\Phi_a$ and $\Phi_p$ are the respective values of the potential, angular
momentum conservation  reads $r_a \cdot v_a = r_p \cdot v_p$ and energy conservation
$v_p^2 - v_a^2 = 2 \cdot (\Phi_a - \Phi_p)$. This gives   
\begin{equation}
 v_p^2 = \frac{v_p^2 - v_a^2}{1 - (r_p/r_a)^2} 
\end{equation}
and

\begin{equation}
v_a^2 = \frac{v_p^2 - v_a^2}{(r_a/r_p)^2 - 1} .
\end{equation}

For a logarithmic potential

\begin{equation}
 v_p^2 - v_a^2 = v_0^2 \cdot \ln\big(\frac{r_0^2 + r_a^2}{r_0^2 + r_p^2}\big). 
\end{equation}

For the above values of $v_0$ and $r_0$, Fig.\ref{orbits} plots perigalactic
velocities vs. pericentric distances for five indicated apocentric 
distances in kpc. Overplotted are all objects with radial velocities lower
than 800 km/s, interpreting their projected distances as pericentric
distances and their radial velocities relative to NGC 1399 as pericentric
velocities. These are the minimal pericentric velocities and the minimal apocentric
distances. One concludes that the radius of the cluster system was  at least
200 kpc, if indeed all clusters were bound to NGC 1399.
 
Some of them are very faint and  we have no independent 
confirmation by line velocities, so doubts may remain. But then also remains 
the question why there are not so many  at the corresponding high
radial velocities. Can they belong to NGC 1399? Relative velocities up
to 200 km/s larger than the circular velocity can be easily produced
by orbits with apocentric distances up to 50 kpc. But for the plotted
objects, one needs distances of 200 kpc and more.
 If by some strange coincidence we would pick up 7-8 clusters with apocentric
distances of 200 kpc, why then only 4-5 with distances of 100 kpc and not,
say, 50, considering the radial power-law number density profile of clusters?

Can the objects with the highest relative, i.e. lowest, radial velocities be
stars? The details of that discussion must be postponed, but an example can
be given. The object 86:105 (in the notation of Paper III) has a radial
velocity of 570 km/s, an R-magnitude of 20.8 and a color of C-R=1.2. The most
favourable case, i.e. the nearest possible distance requires an extremely
metal-poor main sequence star with an absolute magnitude of $M_R = +6$,
corresponding to a distance of about 9 kpc. The distance to the Galactic
center then is roughly 12 kpc. The vector of galactic rotation towards
the Fornax cluster has the value -120 km/s. We thus would have a star at
12 kpc galactocentric distance with a radial velocity of +420 km/s in the
 galactic standard of rest and would face a similar problem as before. 

Given all this, we therefore call attention to the possibility that some
of the objects with radial velocities lower than 800 km/s
are objects in the foreground at an unknown distance and that 
their velocities have to be interpreted as recession rather than peculiar
velocities. 

The existence of a very diluted stellar population between the Local Group
and the Fornax cluster would be an extremely interesting matter for a number
of obvious reasons. HST observations could perhaps resolve nearby objects
into stars. The above cited example should have a distance of only a few
Mpc. If all this was true, one would expect a comparable number per area of
 low velocity objects at radii larger than 10 arcmin away from NGC 1399,
again a strong motivation to observe a distant cluster sample.  

Another possibility is that some objects with extreme velocities are not bound
to NGC 1399. Within a projected radius of 400 kpc, there are more than 20 neighboring
galaxies, making it plausible that an object coming from so far is not necessarily
bound. But then one should expect that there are many more unbound objects with
low inclined orbits.  

 Of course, the mass profile of NGC 1399 has to be derived under the assumption
that indeed all clusters belong to NGC 1399. It is also worth noting that the
pure existence of high velocities relative to NGC 1399 requires the existence of
dark matter.

\subsection{Comments on scenarios for the formation of the cluster system}

In combination with the wide-field Washington study of Dirsch et al. (Paper I)
 we presently have the largest data set available for a
 GCS of a giant elliptical galaxy. Although, as we have seen, such a large
sample produces a lot of new questions, can we formulate constraints for
GCSs formation scenarios currently discussed in the literature? A compact
summary of model predictions is given by Rhode \& Zepf \cite{rhode2001}.
They distinguish four different models: The ''monolithic collapse''
scenario (e.g. Larson \cite{larson1975}, Arimoto \& Yoshii \cite{arimoto1987}), the merger model put forward by Ashman \& Zepf \cite{ashman1992},
and the models by Forbes et al. \cite{forbes1997}, and C\^ot\'e et al. \cite{cote1998}.
The Forbes et al. model let both
metal-rich and metal-poor clusters form during a dissipational collapse and
involves the capture of additional GCs by tidal
effects from neighbouring galaxies or the accretion of dwarf galaxies.
 C\^ot\'e et al.'s model  suggests that only the metal-rich clusters have
formed in a dissipational collapse, while the metal-poor clusters have
been assembled hierarchically by dissipationless mergers.
Rather than to underline the merits of the above suggestions (they are found
in the respective papers) we want to point to a few problems which may arise
when interpreting the properties of the NGC 1399 GCS in the context of
any of these scenarios.

The global orbital distribution of the clusters is close to isotropic.
This requires some kind of relaxation mechanism which works out to large radii. It is
not plausible that a monolithic collapse model (which otherwise has been
successfully applied to explain many properties of elliptical galaxies, e.g. 
Chiosi \& Carraro \cite{chiosi2002} and references therein), 
 can provide such a mechanism. Moreover, in this model, one would expect rotation to be more
pronounced in the inner regions, while there is only a weak signal for
the outer blue clusters.   A further argument from the photometric work is that
the local specific frequency of the clusters is increasing outwards.
 On the other hand, there is evidence
that it depends on the star formation rate (see the remarks below). Therefore, if the star
 formation rate increases inwards according to the gas density
profile, one expects to see the cluster formation efficiency increasing
inwards as well (Larsen \& Richtler \cite{larsen1999},\cite{larsen2000}).
This should not result in an outwards increasing local specific frequency, as is the case in NGC 1399 (Paper I), unless cluster
destruction processes significantly shaped the cluster number density
profile in our radial range. This, however, is unlikely. One then would expect
an increasing tangential bias for both red and blue clusters, for which there
is no evidence.

We note that the remark on the uncertain anisotropy also applies to the analysis of C\^ot\'e et al. \cite{cote2001} of the
M87 cluster system. They find the red clusters to have $\beta = +0.4$ and the
blue ones to have $\beta = -0.4$, when they impose the mass profile taken
from X-ray analyses. While the red clusters probably always stay more radial,
 both the absolute values and the difference are uncertain and may change
 strongly when the mass profile is changed only moderately. A similar approach
in the case of NGC 1399 would have to face the contradicting mass profiles 
and therefore is prohibited, unless there is better agreement.
 
Can the blue clusters be accreted as in the scenario of C\^ot\'e et al. \cite{cote1998}? Invented primarily to explain the phenomenon of bimodality,  we note that one
of its predictions, namely the high specific frequency of central galaxies, is questioned by the
fact that NGC 1399 is probably more ''normal'' regarding its specific frequency than previously thought (Paper I, Ostrov et al. \cite{ostrov1998}). However, C\^ot\'e et al. refer to
M87, where a wide-field study  still has to come. 
Furthermore, C\^ot\'e et al. do
not expect to find a bimodal GC color distribution in dwarf ellipticals, while there is now the example
of NGC 1427, a low luminosity elliptical with bimodality (Forte et al. \cite{forte2001}).
They also cite the near equality of the velocity dispersion of outer globular clusters with
Fornax galaxies found by Kissler-Patig et al. \cite{kissler1999} as evidence for the blue clusters to be tidal debris.
 Our velocity dispersions are indeed similar to that of the
Fornax giant
 ellipticals, but distinctly lower than those of the dwarf galaxies. Drinkwater et al.
 \cite{drink2001} quote $308\pm30$ km/s for giants and $429\pm41$ km/s for dwarfs.

Globular clusters cannot be accreted individually, for example from a free-floating intergalactic population, because neither dynamical friction nor tidal
interaction with the accreting host are sufficient  for those compact and low-mass
objects to create a strongly bound population. 
 They have to be brought in, 
as in the scenario of C\^ot\'e et al., as  
 members of the cluster system of an infalling  galaxy. The question is whether the above processes are able to
produce an isotropic or slightly tangential orbit distribution
of GCs at galactocentric radii between 10 and 40 kpc. Trustworthy answers must
rely on N-body simulations. What we qualitatively can take from such
simulations (e.g. van den Bosch \cite{bosch1999}) is that infalling satellites of sufficient mass on circular
orbits spiral inwards on a time scale of about a few Gyr. But such satellites
(model 6 of van den Bosch et al. has a mass of $2 \cdot 10^{10} M_{\odot}$)
are not tidally destroyed until they reach small perigalactic distances, so
they should still be  visible. Tidal disruption occurs effectively on strongly
elongated orbits (Seguin \& Dupraz  \cite{seguin1996}), where
 circularization of the orbits can occur (Tormen et al. \cite{tormen1998}).
However, clusters tidally released from those satellites and now found again at large
galactocentric distances, must have been released at a very early stage
in order to maintain their highly eccentric orbits.
In total one would expect a radial bias and not isotropy or even a tangential
bias.
Moreover, one would not expect to see any rotation, as possibly seen for
the outer blue clusters.
 It thus seems improbable that a large fraction of the blue clusters
have been accreted through the infall of dwarf galaxies.

A kinematic study of the clusters at larger galactocentric radii may turn out to be more 
 discriminative between the collapse, infall or merger scenarios.
If the rotation of the outer blue clusters is confirmed, an infall scenario becomes less probable,
while rotation can be understood as a relic of a major merger event, as in the case of M87
(Kissler-Patig \& Gebhardt \cite{kige1998}, C\^ot\'e et al. \cite{cote2001}).

\section{Conclusions and Outlook}

We obtained radial velocities for  468 globular clusters  of
the globular cluster system of NGC 1399, the central galaxy in the
Fornax cluster. The clusters have projected radial distances between
2$\arcmin$ and 9$\arcmin$, corresponding to 11 and 50 kpc. The main results emerging from this study are:

We do not find any signature of rotation except for the outer metal-poor
clusters, where a marginal rotation amplitude might be present. 

The velocity distribution of the clusters is not easy to interprete.
 Particularly
under error selection, one finds
an asymmetry with respect to the systemic velocity (1441 km/s). There are  
more objects with extreme low  velocities down to 
600 km/s than at the high velocity wing. A strong error selection produces velocity limits at radial velocities
symmetric to the systemic velocity, suggesting that there are indeed only very few objects with velocities
larger than 2080 km/s or lower than 800 km/s. 
We suggest that a foreground  population of 
sources may contaminate the low velocity sample. Another possibility is that they are not bound to NGC 1399.   

When omitting the extreme velocities, the projected velocity dispersion of the  sample is 274$\pm$9 km/s. It remains
radially constant within the uncertainties. For the non-selected sample, it rises
to 325$\pm$11 km/s, as well radially constant. This constancy stands in contrast
to earlier investigations with smaller samples, which found a strong radial 
increase of the velocity dispersion.

Dividing our sample into red (metal-rich) and blue (metal-poor) clusters,
we find for the red clusters a velocity dispersion of 255$\pm$13 and for
the blue clusters 291$\pm$14, both radially constant. This difference is in agreement with the assumption
of isotropic orbits and the observed surface density distributions.   

We adopted a spherical dynamical model on the basis of the radial Jeans-equation, and tried to
 constrain $\sigma_r$, the radial component of the velocity dispersion. By projecting a model
 globular cluster system, we obtained circular velocity curves, which
 depend on the anisotropy, which
we cannot determine directly. However, for the red clusters the anisotropy
does not influence the circular velocity much. Adopting isotropy,
we derive a radially constant circular  
velocity of 415$\pm$30 km/s.
  
If a $M/L_R$ value of 5.5 for the stellar component is adopted, the dark matter component is well
 described by a logarithmic potential with
the parameters $v_0 = 365\pm6$ km/s and $r_0 = 11.7\pm0.7$ kpc (the small error bars tell us that the shape
of a logarithmic potential is a very good representation. The dominant uncertainty still is the observed
velocity dispersion.) 
Within our radial range, this potential is indistinguishable from a CDM halo of the NFW type with a 
virial mass of $9.7 \cdot 10^{12} M_\odot$ and a concentration parameter $c$=15 (using the definitions of Bullock et al. \cite{bullock2001}).  
Observations of other central galaxies revealed  much shallower density
 profiles. 

Our mass profile disagrees with the X-ray analysis of Paolillo et al. \cite{pao2002}, which results in a mass profile 
increasing with $\sim r^2$ instead of increasing linearly. One might suspect
that the irregular shape of the X-ray halo is somehow related to this
disagreement. It also cannot, although less certain,  duplicate the mass
profile of Ikebe et al. \cite{ikebe1996}, which predicts the opposite, namely
a shallower profile.

Although our sample of radial velocities is the largest so far obtained for any
elliptical galaxy, important questions remain:  
is there really an asymmetry regarding the extreme velocities or would it
disappear in a larger sample? Is there rotation? Out to what radius can 
the constant circular velocity be followed?
The important next step should be to investigate the outer cluster population beyond 40 kpc.  

\section*{Acknowledgments}
Thanks are due to Petra G\"artner for help with the reductions of some masks. 
We are indebted to Ylva Schuberth for many helpful remarks. The comments
of the referee, Dean McLaughlin, were quite essential for improving the first version of the paper. 
T.R. and D.G. acknowledge  ESO visitorships, during which some of the work
has been done. B.D. thanks the Humboldt Science Foundation, Germany, for
financial support. T.R., B.D., D.G., D.M., and L.I. acknowledge support by the FONDAP Center for Astrophysics, Conicyt 15010003.  

{}

\clearpage

\begin{deluxetable}{ccc}
\tabletypesize{\scriptsize}
\tablecaption{This table lists 
the amplitude
and the position angle of  possible rotation (from North past East) for the entire
radial range, for a selection with $r>$5', for the blue clusters with
$r>$6', and for two further selections in the velocity interval 1280 km/s $< v_r <$
1600 km/s. The latter explores possible rotation as the cause for the double-peak around
the systemic velocity. Rotation  around the minor axis marginally is indicated for the outer
blue clusters. Although the double-peak of the entire sample (upper left panel of
Fig.\ref{disptot}) yields a rotation signal, it vanishes for the red clusters 
in spite of being more pronounced than for the blue clusters (left panel of
 Fig.\ref{redblue}).}  
\tablewidth{0pt}
\tablehead{
        \colhead{} &
        \colhead{A [km/s]} &
        \colhead{$\Theta_0 [degree] $}}

\startdata
all & $10\pm17$ & $53\pm110$ \\
blue & $15\pm26$ & $250\pm102$ \\ 
red & $7\pm24$ & $16\pm197$ \\
\hline
all (r$>$5$\arcmin$) & $26\pm30$ & $125\pm56$\\
blue (r$>$5\') & $22\pm40$ & $150\pm94$\\
red (r$>$5\') & $14\pm54$ & $253\pm270$\\
\hline
blue (r$>$6\') & $68\pm60$ & $140\pm39$\\
1280-1600 (all) & $20\pm8$ & $198\pm27$\\
1280-1600 (red) & $5\pm12$ & $269\pm133$\\
\enddata
\label{rot.tab}
\end{deluxetable}

\begin{deluxetable}{ccccccccc}
\tabletypesize{\scriptsize}
\tablecaption{This table lists the results of the (projected) velocity dispersion measurements. The first column
indicates the radial bin, the second the dispersion of the entire sample, followed by the number of objects (third
column). The fourth column gives the dispersion for the blue clusters followed by the number of objects (fifth column). The
sixth column gives the dispersion for the red clusters, followed by the number of objects. The uncertainties result
from the expressions given by Pryor \& Meylan \cite{pryor1993}. See the text for further comments.}

\tablewidth{0pt}
\tablehead{
        \colhead{} &
        \multicolumn{2}{c}{no selection} &
        \multicolumn{2}{c}{all selected clusters} &
        \multicolumn{2}{c}{blue clusters} &
        \multicolumn{2}{c}{red clusters}\\ 
      
        \colhead{$\Delta r$ [arcmin]} &
        \colhead{$\sigma$[km/s]} &
        \colhead{N} &
        \colhead{$\sigma$ [km/s]} &
        \colhead{N} &
        \colhead{$\sigma$[km/s]} &
        \colhead{N} 
}
\startdata
all r & 325$\pm$11 & 468 & 274 $\pm$ 9 & 444 &  291 $\pm$ 14 & 213 &  255 $\pm$ 13  & 213 \\
 r$<$3.5 & 322 $\pm$ 18 &  158&  256 $\pm$ 15 & 148 &  267 $\pm$ 23 & 71 &  246 $\pm$ 20  & 77 \\
2.5 $<$r$<$5.5 & 342 $\pm$ 16 & 231 & 285 $\pm$ 14 & 219 &  307 $\pm$ 21 & 109 &  261 $\pm$ 18  & 110 \\
4.5 $<$r$<$7.5 & 306 $\pm$ 14 & 234& 267 $\pm$ 13 & 206 &  281 $\pm$ 19 & 114 &  249 $\pm$ 18  & 92 \\
 r$>$6.5 & 316 $\pm$ 24 & 91 & 280 $\pm$ 22 & 86 &  297 $\pm$ 33 & 47 &  258 $\pm$ 31  & 39 \\

\enddata
\label{dispersions}
\end{deluxetable}

\begin{deluxetable}{cccccc}
\tabletypesize{\scriptsize}
\tablecaption{This table lists
the coefficients in the adopted linear dependence of $\sigma_r$ on galactocentric 
distance ($\sigma_r = a_1 \cdot r + a_2$), which give radially constant projected
velocity dispersions for  different assumed anisotropies for blue/red clusters ($\sigma_r$
in km/s, r in kpc).
 }
\tablewidth{0pt}
\tablehead{
        \colhead{} &
        \multicolumn{5}{c}{blue clusters/red clusters}\\ 

        \colhead{coeffs} &
        \colhead{$\beta$ = 0.8} &
        \colhead{$\beta$ = 0.4} &
        \colhead{$\beta$ = 0.0} &
        \colhead{$\beta$ = -1.0} &
        \colhead{$\beta$ = -5.0}\\ 
}
\startdata
$a_1$ &  -\,/0.4 & 0.43/0.1 &  0/0  & -0.35/-0.2 & -0.45/-0.22 \\
$a_2$ &  -\,/360 & 315/294 &  291/255  & 245/205 & 164/130 \\
\enddata
\label{coeffs}
\end{deluxetable}

\clearpage

\figcaption[richtler.fig1.ps]{
This plot shows the relation between the R-magnitude of the clusters and
the uncertainties of their radial velocities (filled circles). For comparison
 the
magnitudes and velocity uncertainties
for the sample of M87 clusters of Hanes et al. \cite{hanes2001}, for
which errors below 200 km/s are quoted, are also plotted (squares).
\label{errorplot}}

\figcaption[richtler.fig2.ps]{
This plot shows the distribution of clusters on the sky for
which correlation
velocities are available. Note that the x-axis is right ascension, and one has to multiply
it with the cosine of the declination to get the same projected scale as the y-axis.
\label{distribution}}

\figcaption[richtler.fig3.ps]{
This plot shows the color-magnitude diagram  for
the sample of Fig.\ref{distribution}. Plotted is C-R (Washington C, Kron R)
vs. R. The bimodal color distribution of the NGC 1399 cluster system
is more pronounced  in a larger and fainter sample (see Paper I) except for the
bright clusters, which do not show a bimodal distribution.
We define the limit distinguishing red and blue clusters to be C-R = 1.6.
\label{CMD}}

\figcaption[richtler.fig4.ps]{
Velocity histograms of the entire sample (upper left panel), an error
selected sample (upper right panel), an inner sample
(lower left panel), and an outer sample (lower right panel). The bin size is
70 km/s. The vertical line at 1441 km/s indicates the systemic velocity. The solid lines
 indicate the cross-correlation
velocities, the dashed lines the velocities through line measurements.
In the upper panels, Gaussian fits with a dispersion of 290 km/s are overplotted.
 Note the
seemingly  non-Gaussian
appearance due to the double peak and the apparently asymmetric velocity distribution. 
A possibility is that the  two peaks are caused by rotation
of a subsample, but no support has been found for that.    
The peak at 1800 km/s is also apparent in the sample of Kissler-Patig et al. 
\cite{kissler1999}
\label{disptot}}

\figcaption[richtler.fig5.ps]{
This plot shows six simulated Gaussian distributions, each containing 470 objects.
The numbers are the probabilities of being drawn from Gaussians according to
 KS-tests.
It illustrates that low probabilities frequently occur. Also peaks or
apparent distortions do not allow conclusions regarding the Gaussian or
non-Gaussian nature. 
\label{gaussim}}

\figcaption[richtler.fig6.ps]{
The upper panel shows the velocities vs.
projected galactocentric distance for all clusters. The dashed -
dotted line is the systemic radial velocity of NGC 1399. In the middle panel,
we selected those clusters for which the difference between correlation
velocity and velocities measured by lines amounts to less than 100 km/s.
The lower panel shows those objects for which this difference is less than
50 km/s. 
The upper limit at about 2000 km/s is already discernable in the unselected sample. 
The symmetric low velocity limit at about 800 km/s shows up only after a strict error selection. 
\label{velorad}}

\figcaption[richtler.fig7.ps]{
Plotted are the R-magnitudes vs. the radial velocities. This sample is somewhat
smaller because there is no photometry available for about 30 clusters.
The vertical lines are
at 2000 km/s (to roughly indicate a velocity, above which only faint objects
are found), at 1440 km/s (the mean velocity of the entire sample), and at 880 km/s,
the corresponding velocity on the low velocity side. It is striking that there are
no bright objects at high velocities, while there are some at very low velocities
(the faint objects may have individually uncertain velocities).
This may indicate that some of the bright clusters actually are located in the foregound.
\label{velomag}}

\figcaption[richtler.fig8.ps]{
This plot shows the velocity distribution (correlation velocities)
for the red and the blue clusters separately. The lower velocity
dispersion of the red clusters is discernible.
Both
distributions show the extended wing towards low velocities. The systemic
velocity is indicated by the dotted vertical line.
\label{redblue}}

\figcaption[richtler.fig9.ps]{
Radial velocities vs. galactocentric distance for blue and red clusters. 
The lower velocity dispersion of the red clusters is visible. 
\label{veloradredblue}}

\figcaption[richtler.fig10.ps]{
This figure is principally suitable to detect
rotation. Plotted are the radial velocities vs. the azimuth angle (East past
North) for the full sample (uppermost panel), the red clusters (second panel),
the blue clusters (third panel). There is no significant rotation present in
either sample.
\label{rotrev}}

\figcaption[richtler.fig11.ps]{
The upper panel shows a selection of velocities between 1280 km/s and 1600 km/s in order
to search for a rotation signal motivated by the double peak in Fig.\ref{disptot}. A rotation
 signal is marginally  present. The middle panel shows the same for the red clusters only because of
 the more prominent double peak but here no rotation is visible. The lower panel selects the blue
clusters more distant than 6 arcmin. This sample again shows marginal rotation.
\label{rotrot}}

\figcaption[richtler.fig12.ps]{
This plot shows the xy-distribution (arcmin) for two samples referring
to the two peaks near the systemic velocity. Triangles are velocities
between 1280 km/s and 1400 km/s. Hexagons are velocities between 1480 km/s
and 1600 km/s. Blue clusters are additionally marked with crosses. See the
text for further comments.
\label{xyplot}}

\figcaption[richtler.fig13.ps]{Shown is the effect of the velocity limits on
the velocity dispersion for the red, the entire, and the blue sample. The
x-axis defines the low velocity cut-off (the high velocity cut-off is always
symmetric with respect to the systemic velocity).
\label{dispvar}}

\figcaption[richtler.fig14.ps]{
This plot visualizes Table \ref{dispersions}. Plotted are the
 galactocentric projected distances vs. the projected velocity dispersions for 
the total unselected sample (uppermost panel) and three velocity selected samples
(800 km/s$<v_{rad} <$ 2080 km/s) in four radial bins: all clusters, 
blue clusters, and red clusters. The dashed
 horizontal lines indicate the velocity dispersions of the respective samples
 for the entire radial range. The radial behaviour is consistent with a
constant projected velocity dispersion. For reasons explained in the text,
we consider the velocity selected sample for further analysis.
\label{dispdat}}

\figcaption[richtler.fig15.ps]{
This plot is the result of projecting a model cluster system under the condition that
the projected velocity dispersion is constant with radius and reproduces the observed velocity
dispersions for red and blue clusters. Plotted are the resulting circular velocities vs.
distance for a variety of  anisotropy values of the cluster system. The dotted
 line comes from the luminous matter only. The long dashed curve is the sum of the luminous
mass distribution and our logarithmic potential. That the degeneracy of the red clusters
 is modest compared with the blue clusters, is a consequence of their steeper
number density profile in combination with a radially constant anisotropy.
\label{dispproj}}

\figcaption[richtler.fig16.ps]{
This figure compares our circular velocities (assuming isotropy for the red
cluster population) with the radial extrapolation of analyses performed for
the stellar body of NGC 1399. The dashed lines embrace  the acceptable
models of Kronawitter et al. \cite{krona2000}.  The dotted line is the best
model of Saglia et al. \cite{saglia2000}. Note, that other models of
Saglia et al., for which the halo parameters are not given, fit better.
The solid line represents the circular
velocities of the red cluster population assuming isotropy. The crosses mark
the best solution of Kronawitter et al., read off from their circular
velocity diagram.   
The dashed-dotted line indicates the circular velocity which would result from
the full cluster sample without any velocity selection.
\label{dispproj2}}

\figcaption[richtler.fig17.ps]{
This figure illustrates that a CDM halo also can fit the
circular velocities derived from GCs. The lower dashed-dotted line corresponds to a CDM halo
of the type described by Navarro et al. \cite{navarro1997}, while the upper dashed line adds
to that the luminous component (dotted line). 
The CDM halo  has a virial mass of
$9.7 \cdot 10^{12} M_{\odot}$ and a concentration parameter c of
 15. The solid
line corresponds to the sum of luminous matter (dotted line) and this halo. For
comparison, a logarithmic potential with $v_0 = 365$ km/s and $r_0 = 11.7$ kpc
is plotted as well (dashed-dotted). They are indistinguishable except for
the very inner region. The crosses mark the best fit of Kronawitter et al.
\cite{krona2000}.
\label{CDM}}

\figcaption[richtler.fig18.ps]{
A comparison of different mass profiles of NGC 1399. The solid line is the
sum of the luminous mass (lower dotted line) and a logarithmic potential with $v_0$ = 
365 km/s and $r_0$ = 11.7 kpc, corresponding to a constant circular velocity
of 415 km/s. Two dotted lines indicate the mass profile, if the circular velocity
was 450 km/s or 390 km/s, respectively.  
 The dashed-dotted line is the mass derived from ASCA X-ray data according to
Ikebe et al. \cite{ikebe1996}, which is distinctly flatter. 
The long-dashed line is the mass derived from
ROSAT data according to Jones et al. \cite{jones1997}.  The short-dashed line
is the mass profile acccording to Paolillo et al. \cite{pao2002} (see text for
more details), derived from ROSAT data as well. It is apparent that our mass profile does not
 agree neither with Paolillo et al. nor with Ikebe et al.. 
The dash-dot-dot line is a mass profile resulting from the entire sample without any velocity selection,
corresponding to a circular velocity of 465 km/s.
\label{massprof}}

\figcaption[richtler.fig19.ps]{
In this plot, we extrapolate our mass profiles from the inner region out to a
radial distance of 500 kpc in order to see, whether such extrapolation would agree with other 
mass profiles, derived from galaxies or X-rays (the shape of the X-ray profiles are approximate only, 
since they had to be read off from log-log plots). The labels are: 1 - ROSAT (Paolillo et al. \cite{pao2002}),
2 - Drinkwater et al. \cite{drink2001} (the dotted lines mark their confidence limits), 3 - ASCA profile (Ikebe et al. \cite{ikebe1996}), 4 - logarithmic potential with $v_0$ = 365 km/s and $r_0$ = 11.7 km/s, 5 - NFW profile
with $r_{vir}$ = 523 kpc and $c$ = 15. A comparison with the Fornax cluster mass profile
from Drinkwater et al. \cite{drink2001} reveals that neither the NFW profile nor the
logarithmic potential can account for the mass of the inner cluster region.
The ASCA profile can, but we do not duplicate its behaviour in the inner region.
\label{masslarge}}

\figcaption[richtler.fig20.ps]{
This plot shows the pericenter velocities vs. the pericenter distances for bound
objects with various apocentric distances (labelled in kpc), which are moving
in a logarithmic potential, characterized by $v_0$ = 400 km/s and $r_0$ = 0.5
 kpc. This potential is a good representation of the total mass within 40 kpc
galactocentric distance and it is assumed that it may be extrapolated out to
400 kpc. Plotted are objects of NGC 1399 with radial velocities smaller than
800 km/s, interpreting their velocities relative to NGC 1399 as their 
space velocities at their pericentric positions. Already under these minimal
assumptions, the majority of them must have their apocentric points at very
large distances. 
The other possibilities are that they
are not bound or that their radial velocities are recession velocities, i.e.
they are in the foreground.
\label{orbits}}

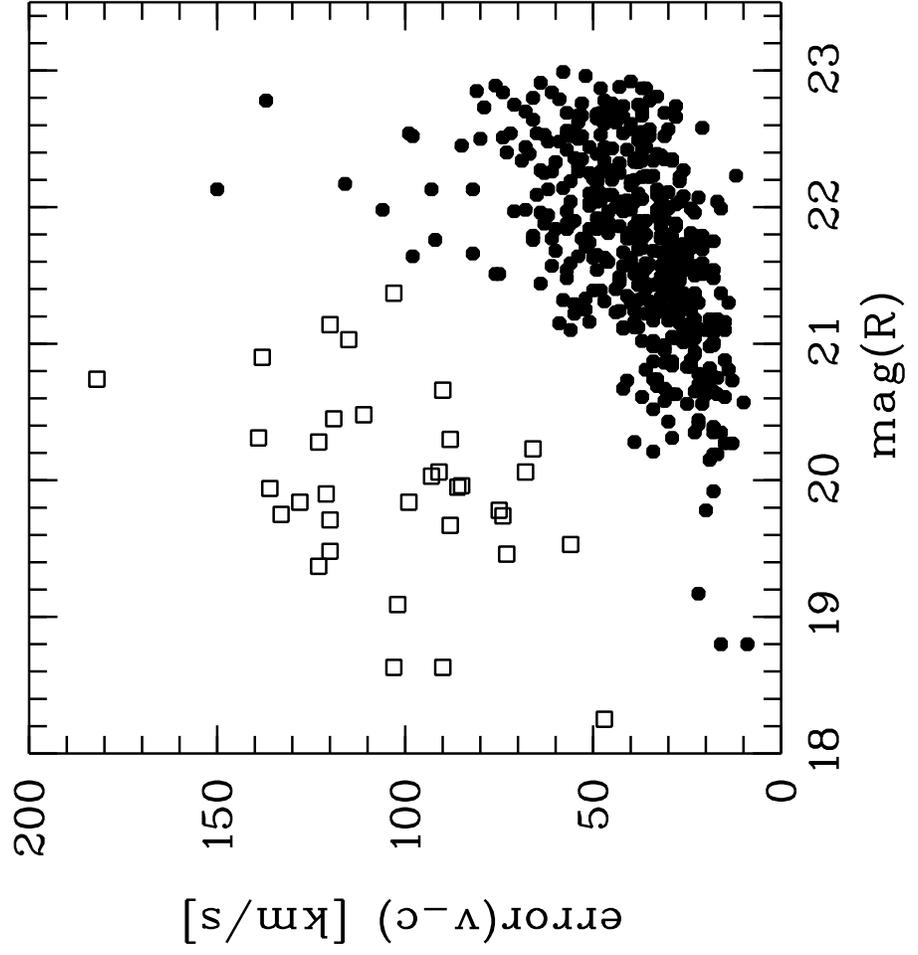
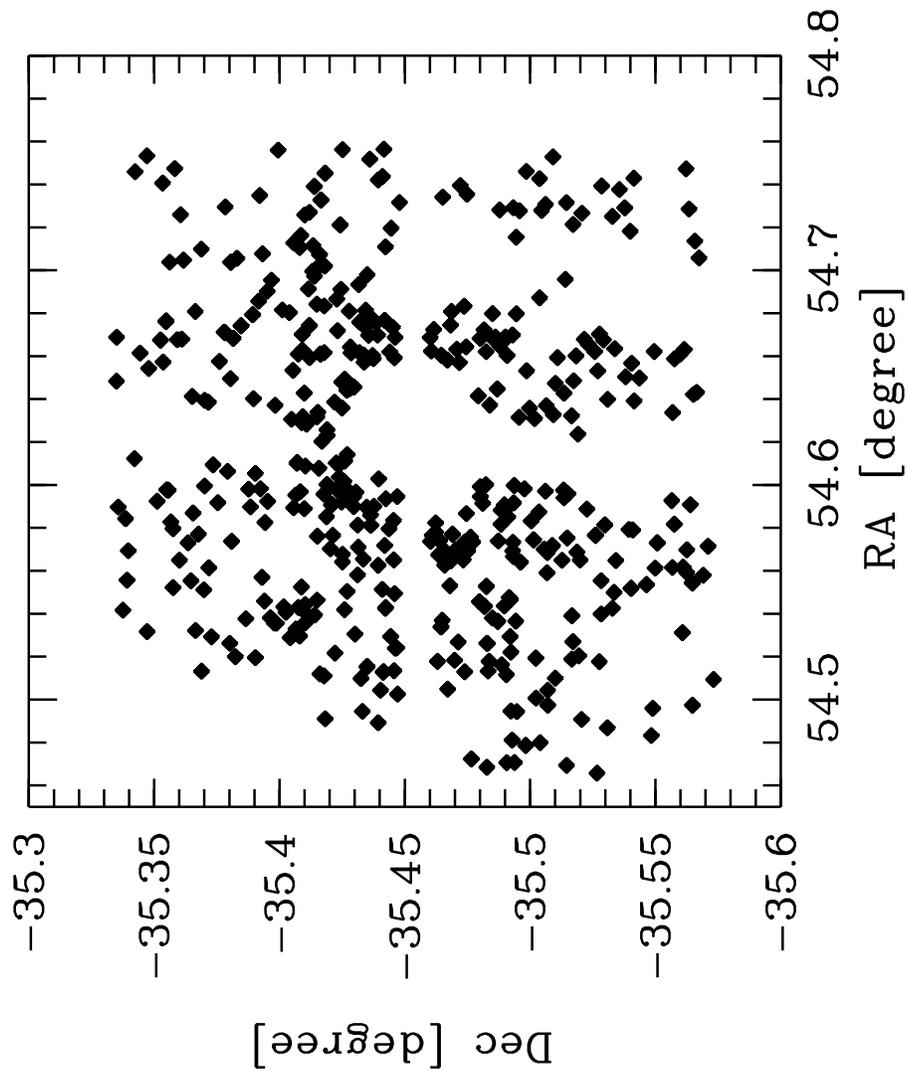
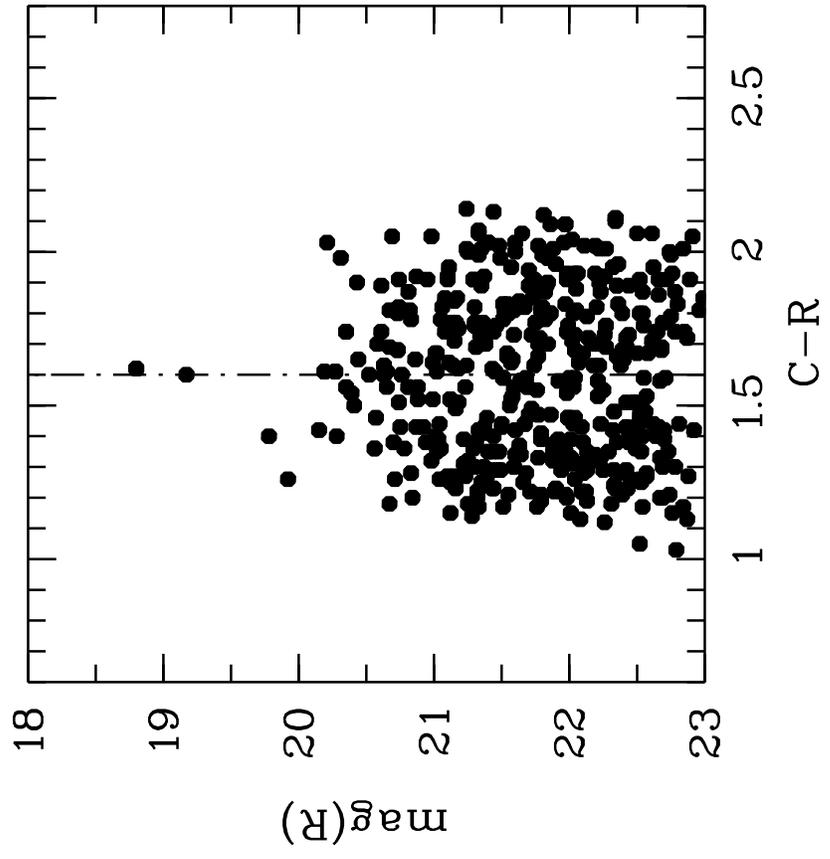
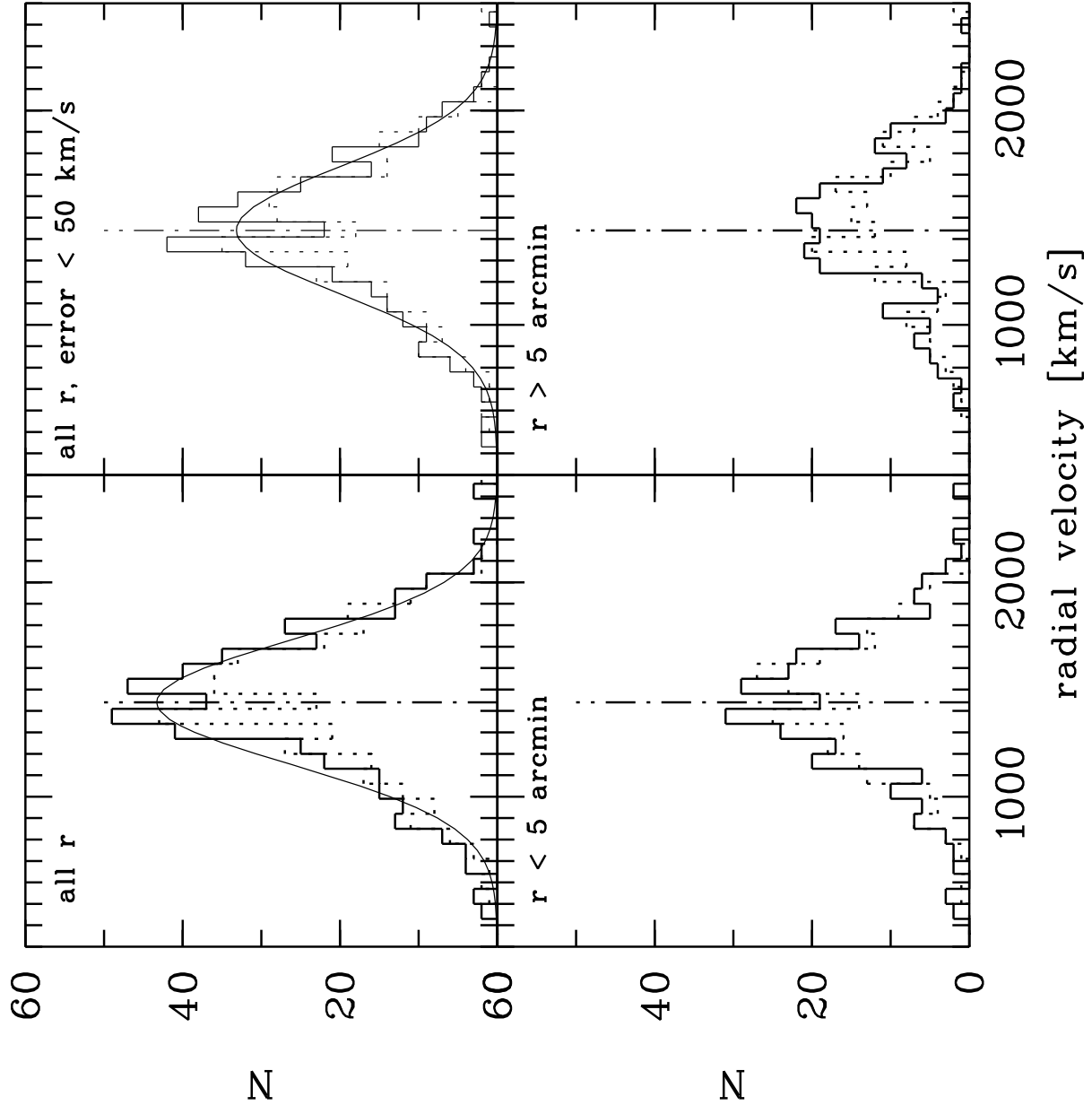
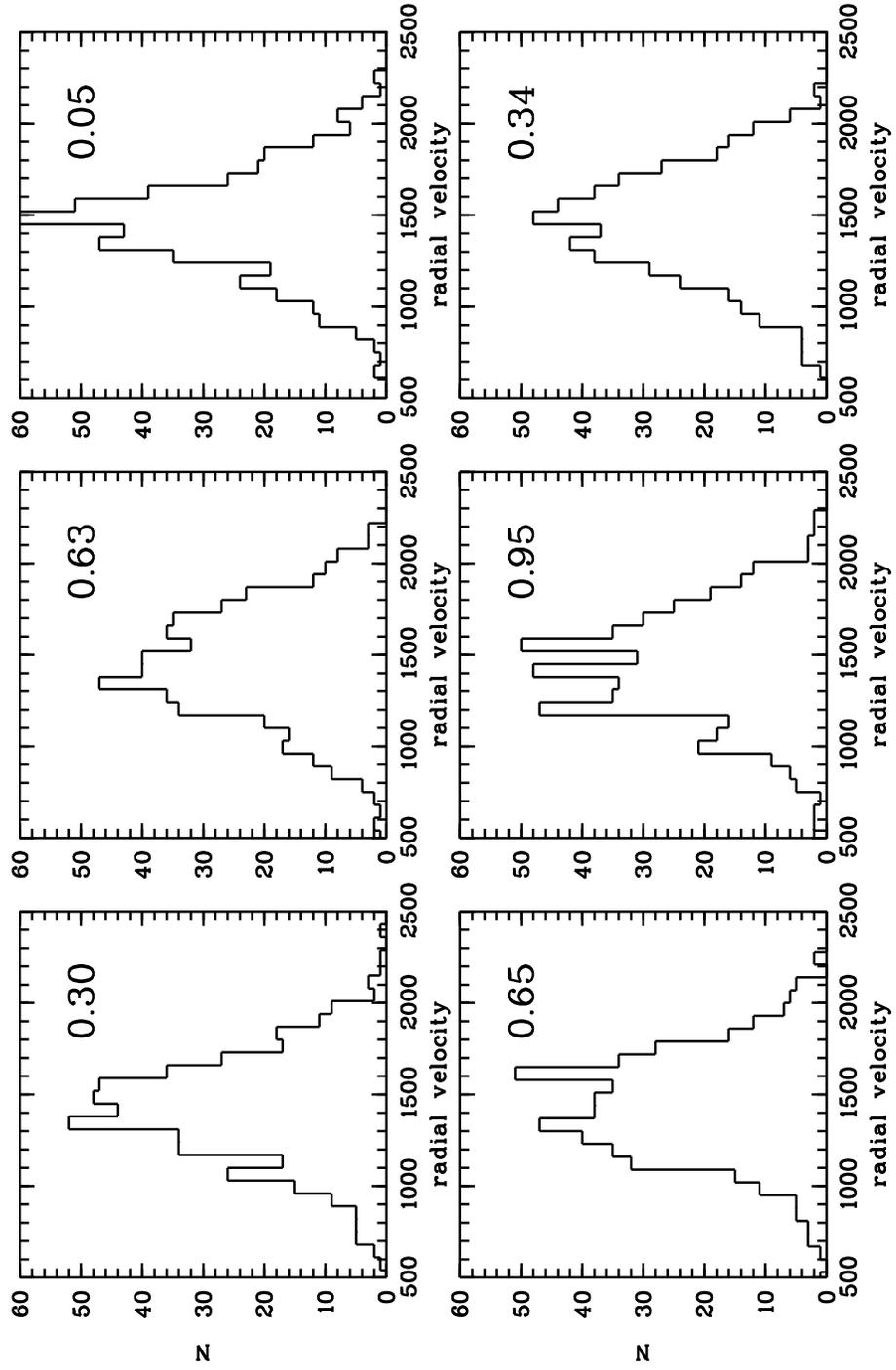
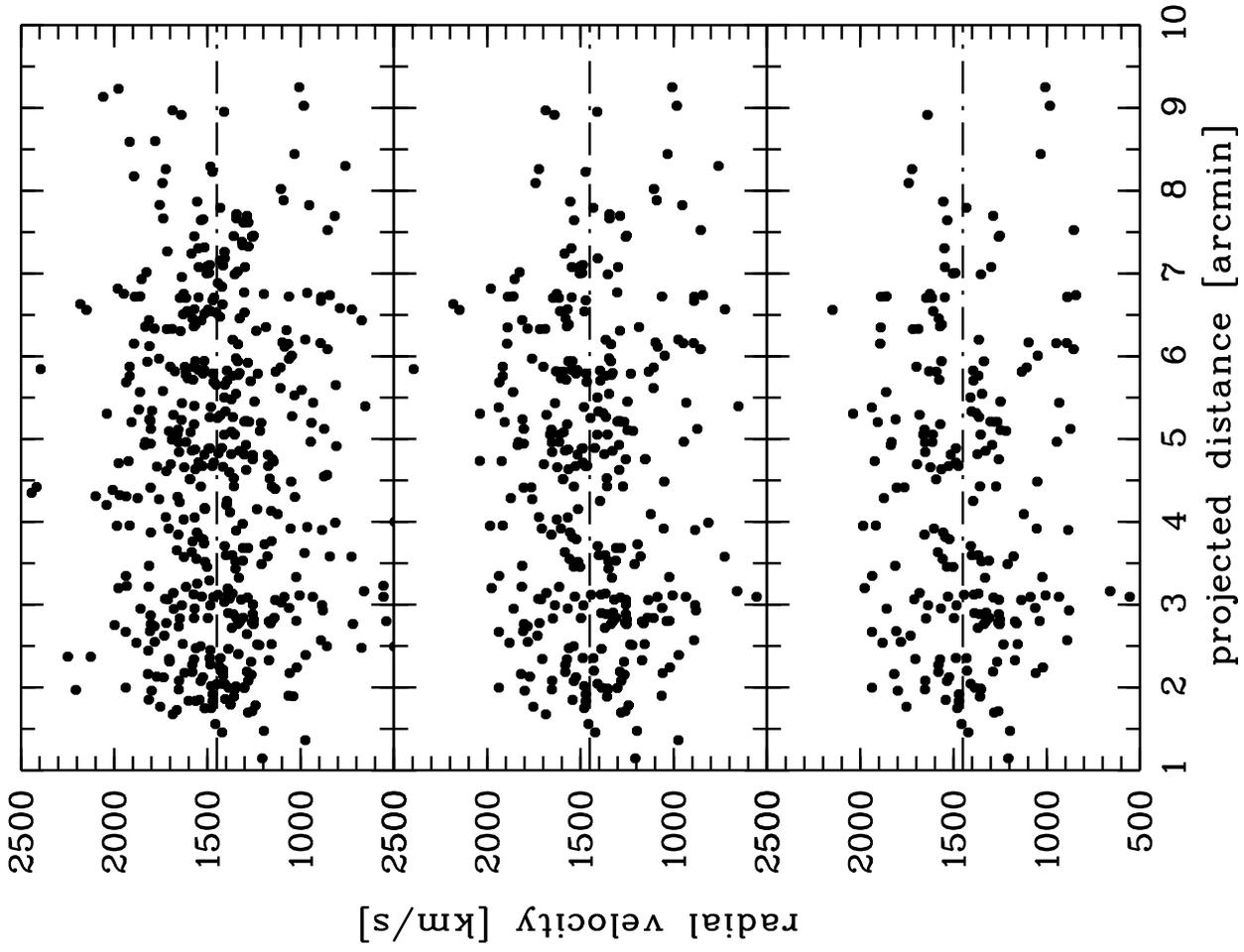
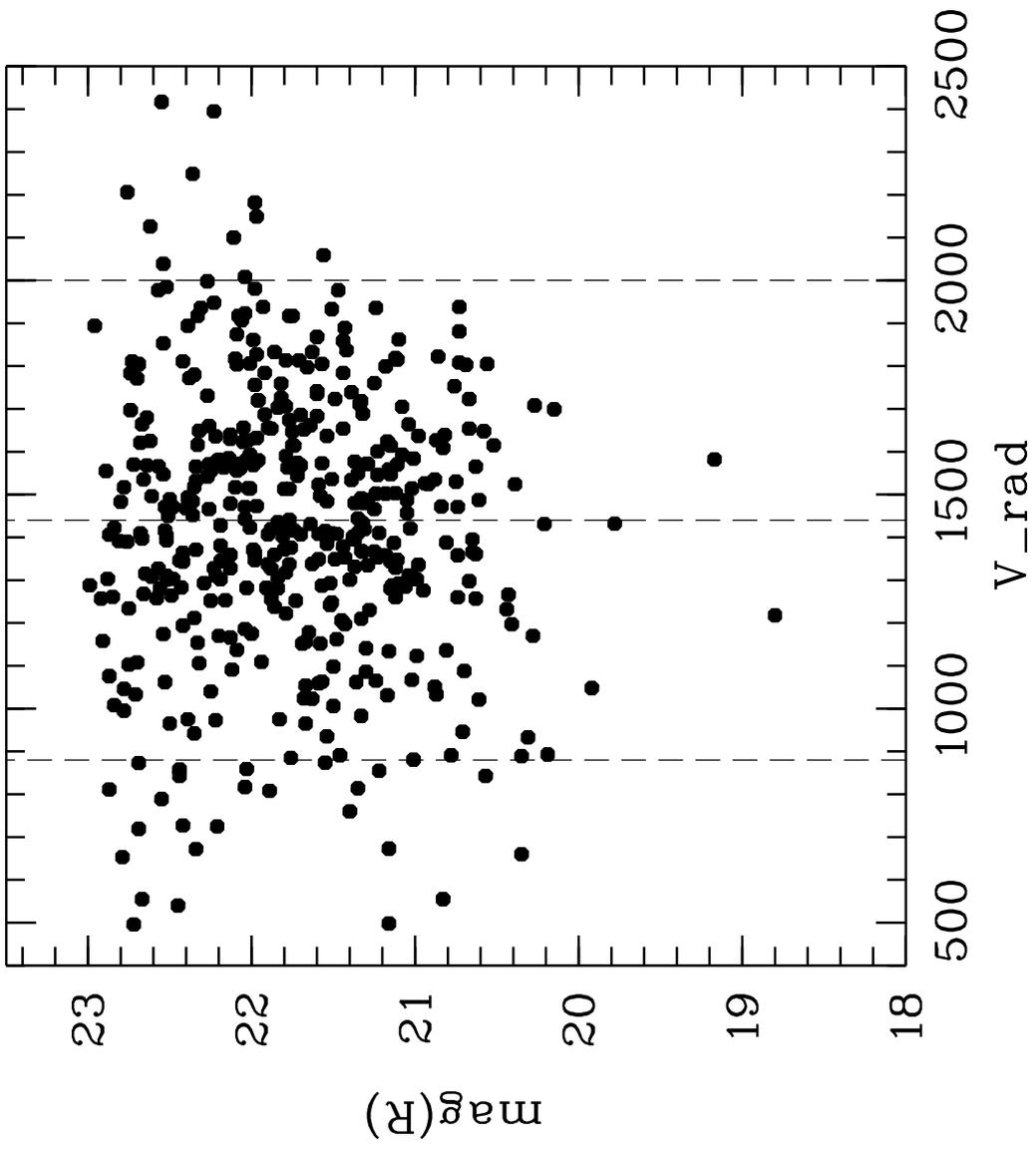
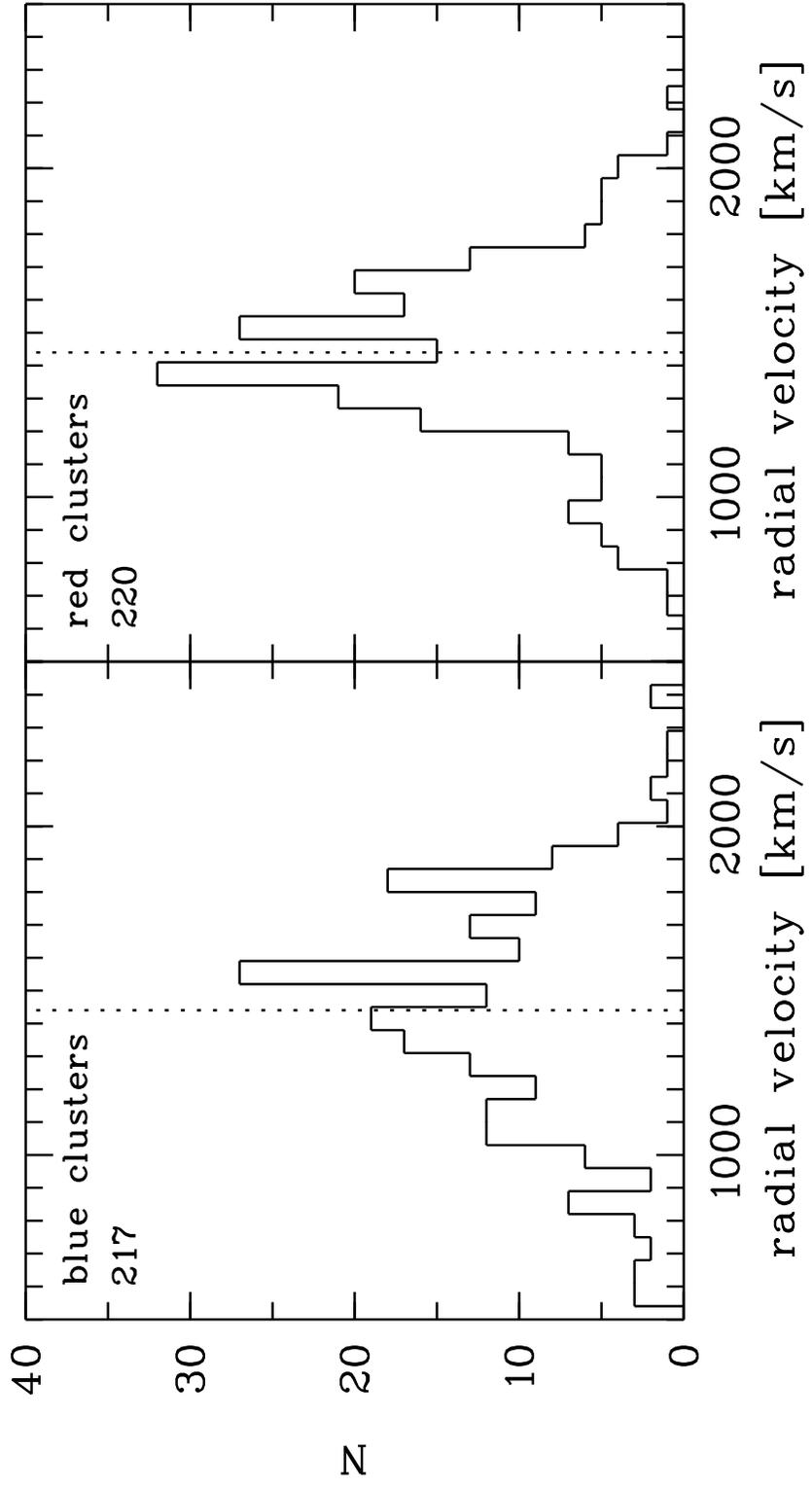
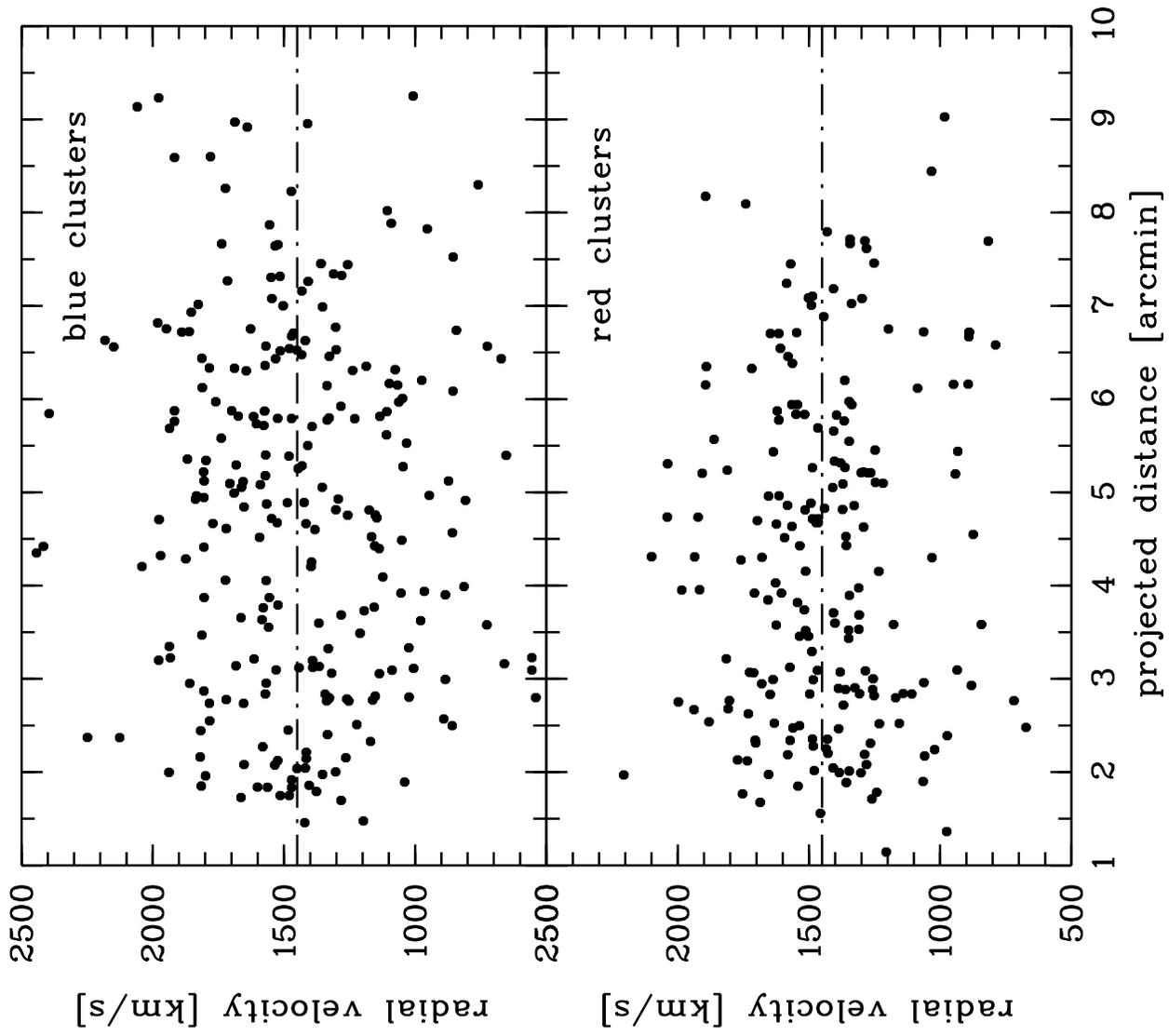
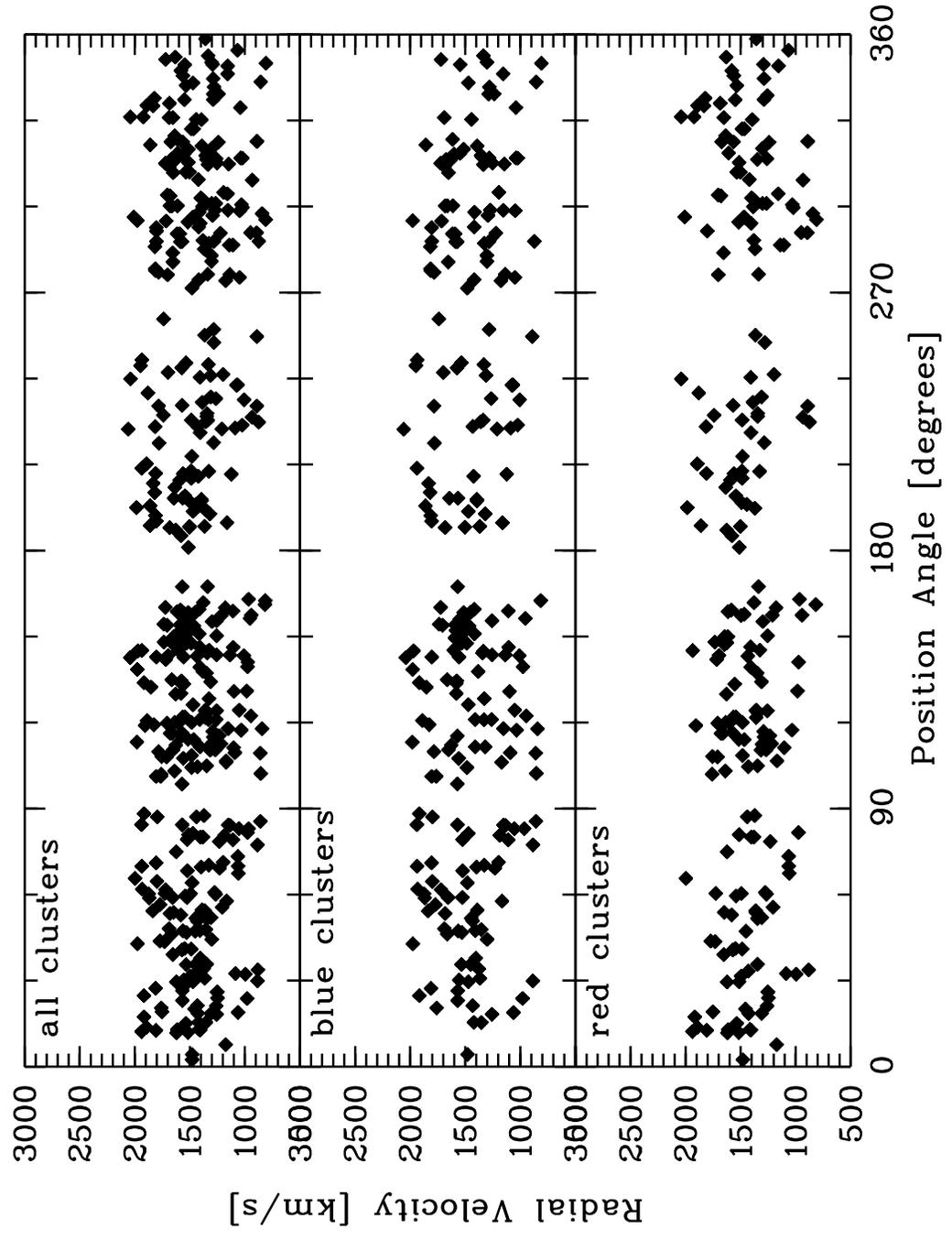
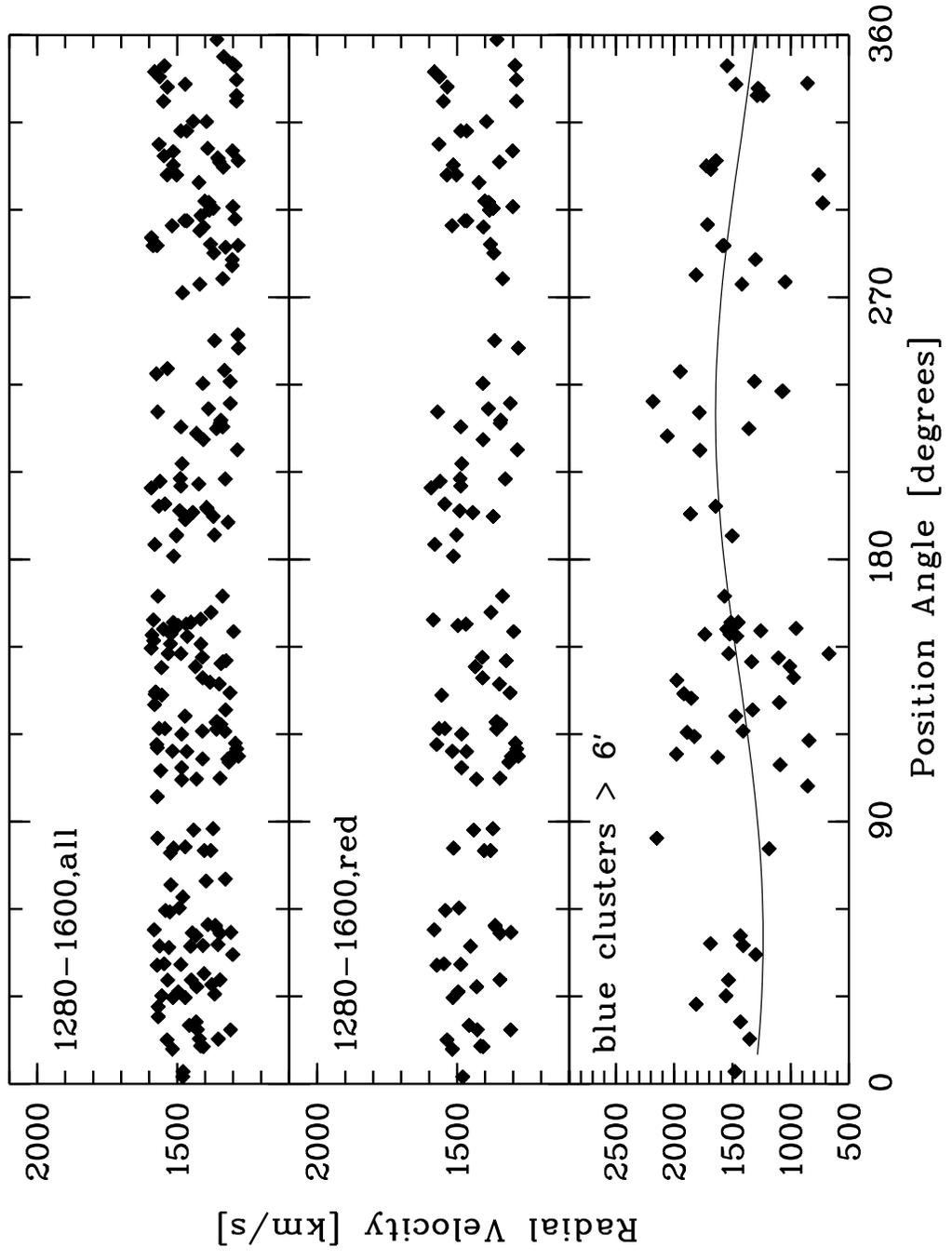
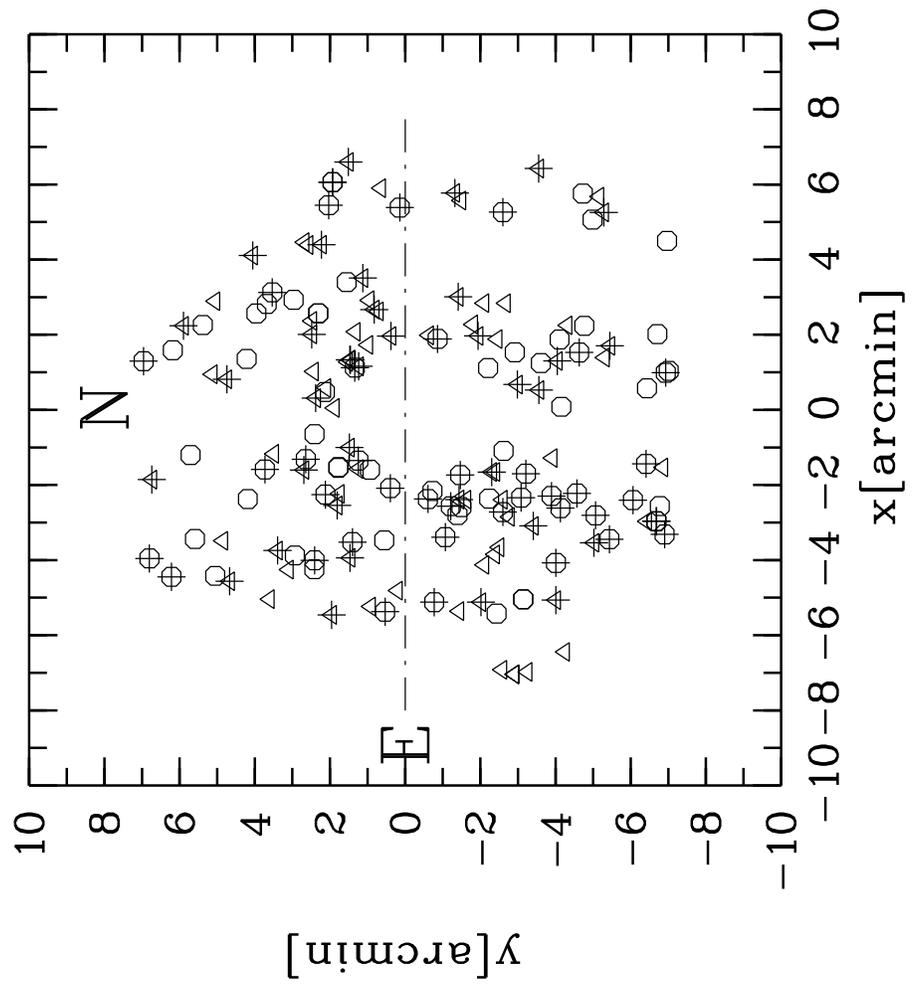
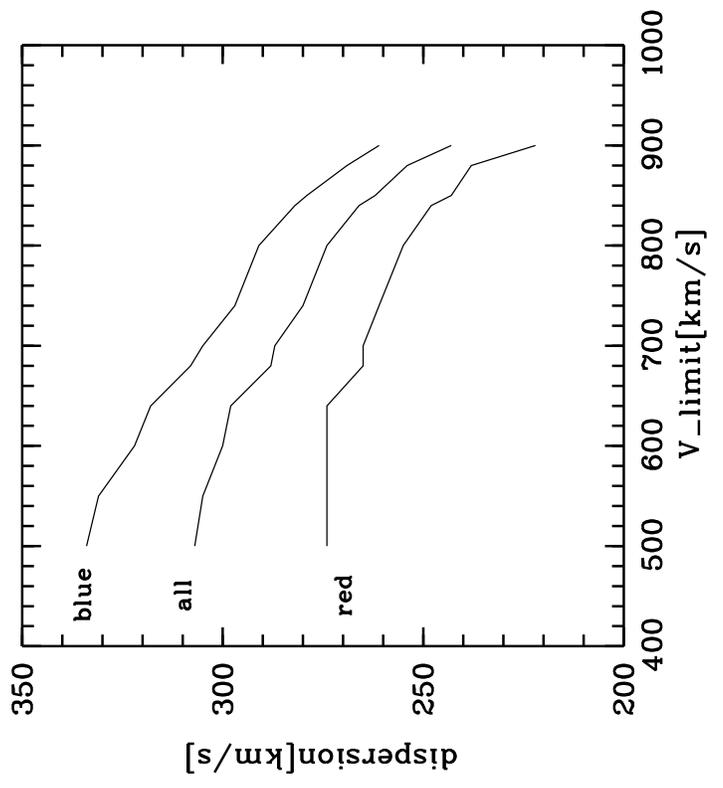
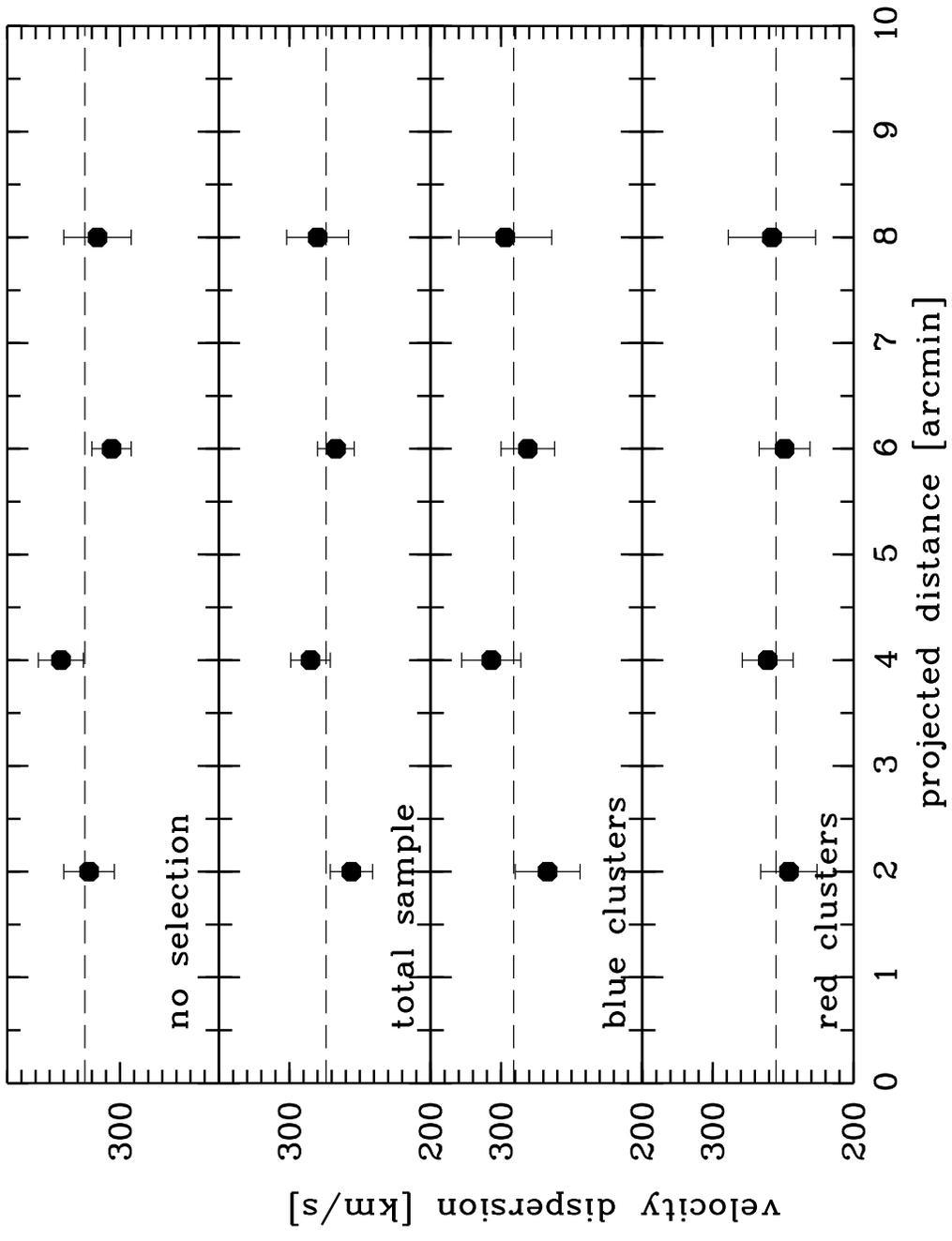
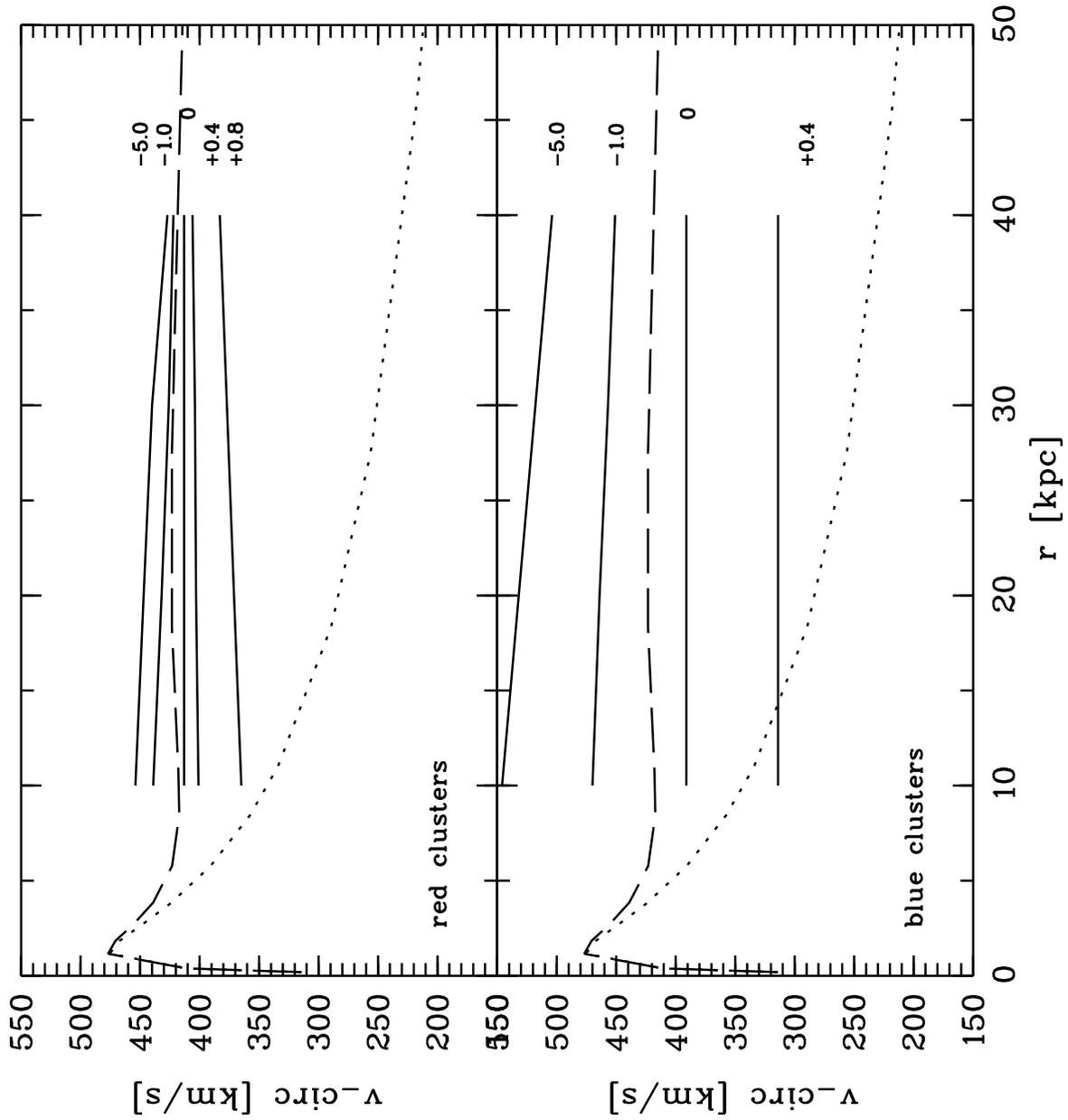
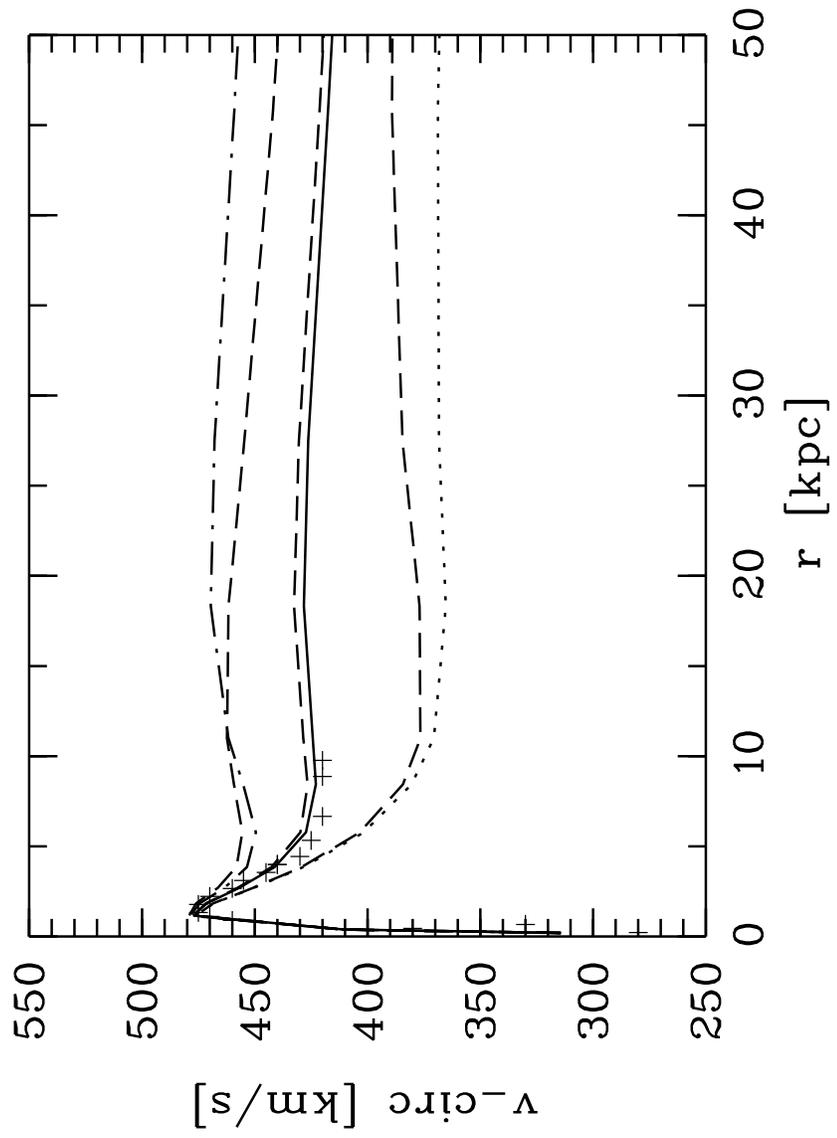
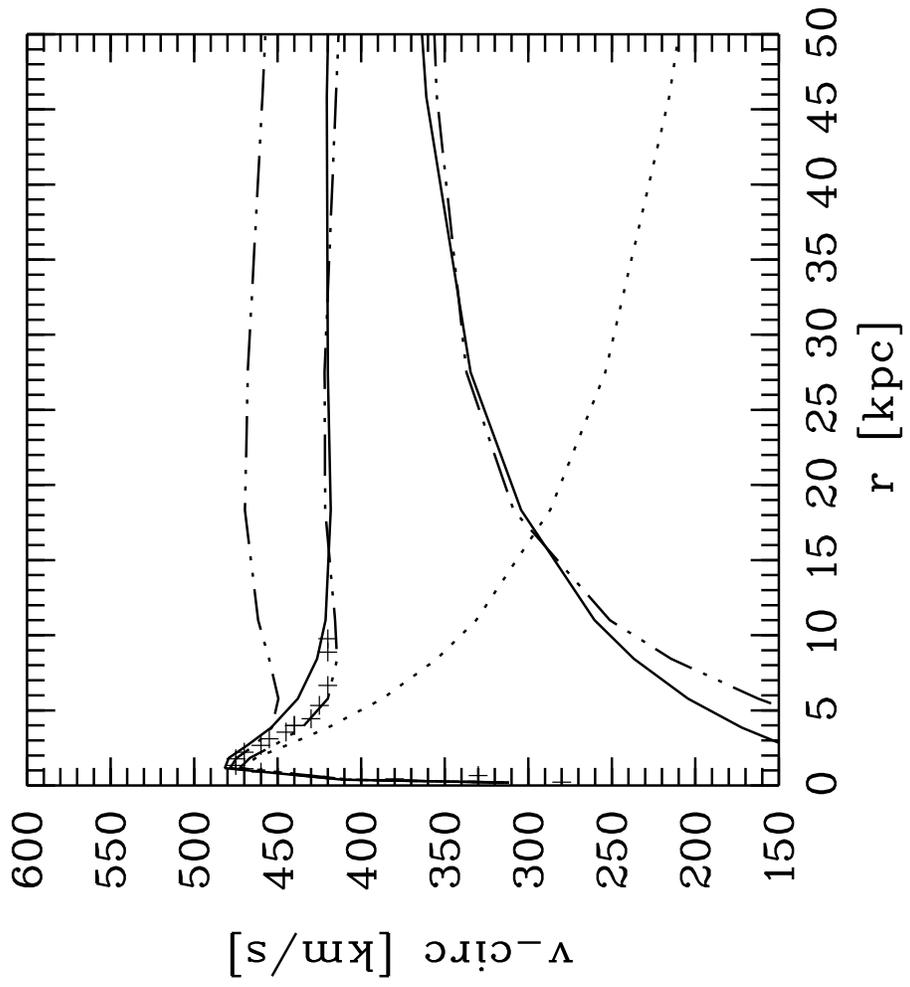
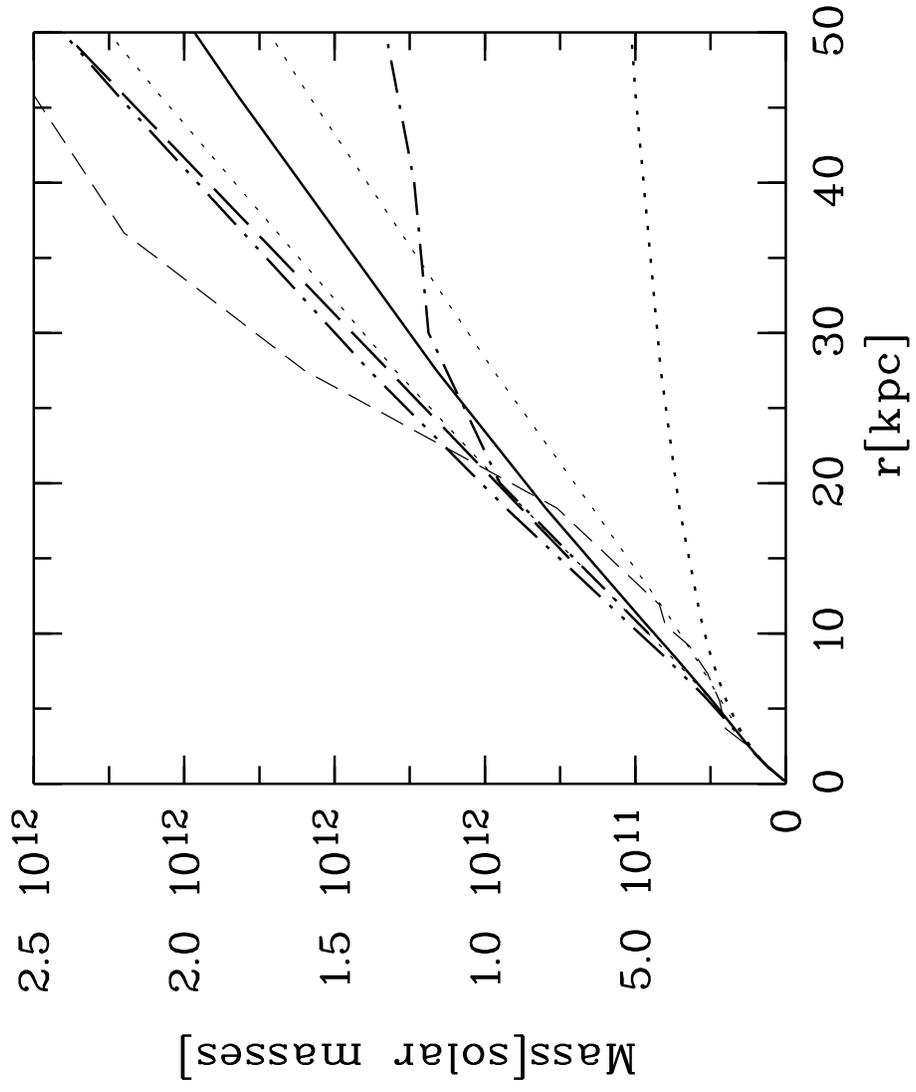
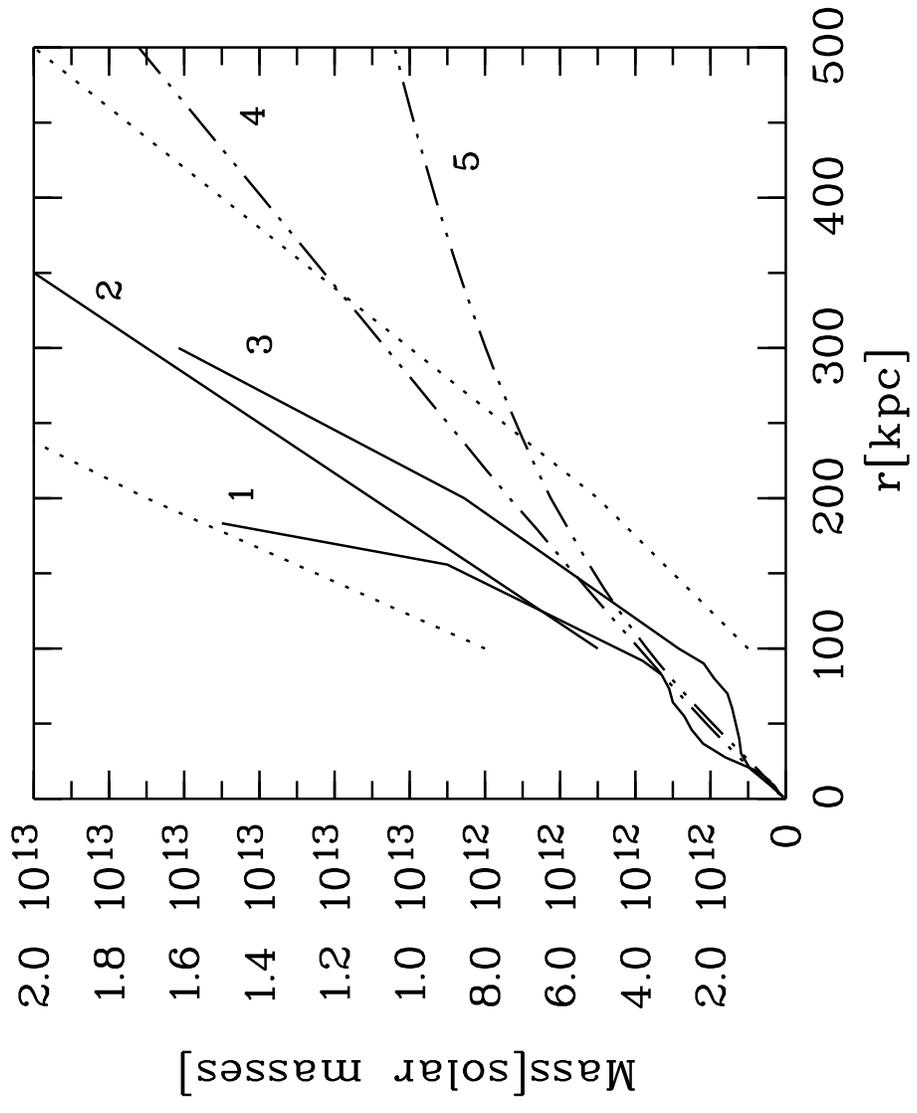
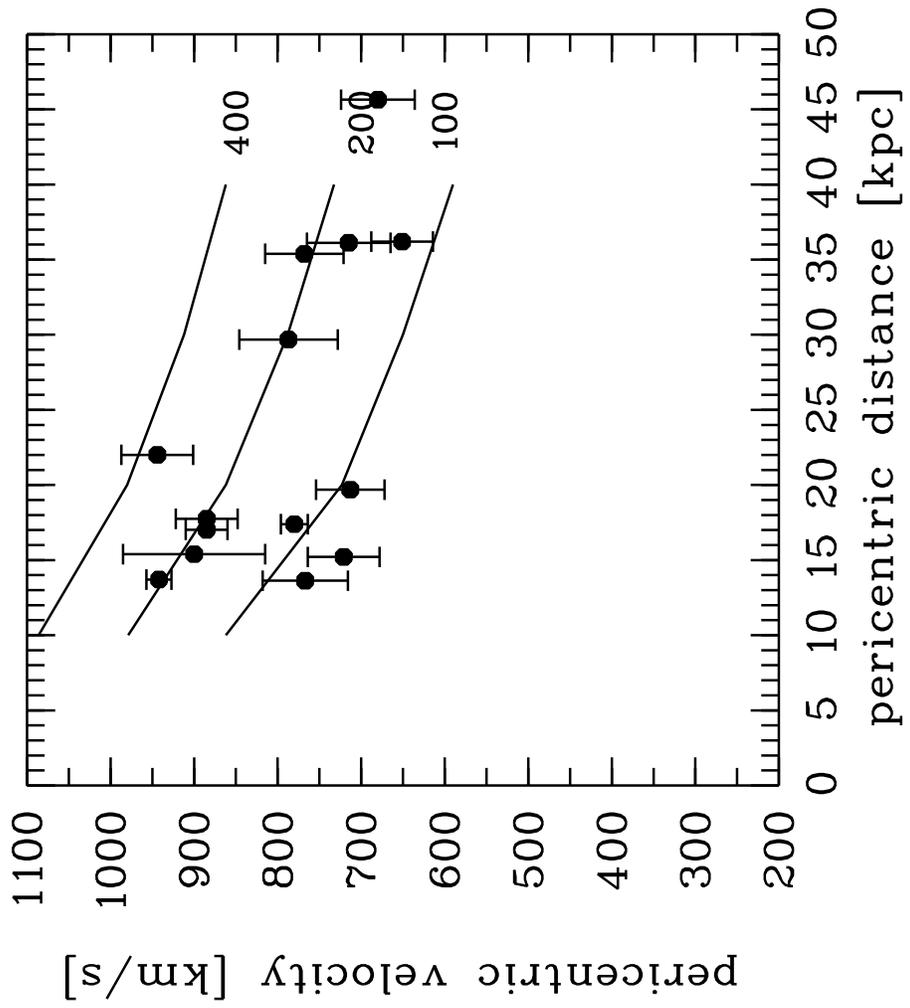
\end{document}